\documentclass[journal]{IEEEtran}

\usepackage{amsmath, bm, hyperref, authblk, nomencl, lipsum, chngcntr, blindtext}
\usepackage{tikz}
\usepackage[utf8]{inputenc}
\usepackage{pbox}
\usepackage{cite}
\usepackage[english]{babel}

\usepackage{array,colortbl,multirow,multicol,booktabs,ctable}

\usepackage{placeins}
\usepackage{forest}
\usetikzlibrary{arrows,decorations.pathmorphing,backgrounds,fit,positioning,shapes.symbols,chains, shapes,decorations}
\usepackage{csquotes}
\usepackage{comment}
\usepackage{color}
\usepackage{url}
\usepackage{amsfonts}
\usepackage{array}
\usetikzlibrary{arrows.meta}
\usetikzlibrary{shapes,arrows}
\usetikzlibrary{automata, positioning}
\usepackage{subfig}

\usepackage[makeroom]{cancel}
\usetikzlibrary{arrows,decorations.pathmorphing,backgrounds,fit,positioning,shapes.symbols,chains, shapes,decorations}

{\title{A Hierarchical Approach to Multi-Energy Demand Response: From Electricity to Multi-Energy Applications}}
\author{Ali Hassan, Samrat Acharya, Michael Chertkov, Deepjyoti Deka and Yury Dvorkin}	
\date{} 
\usepackage{url}
\usepackage{amsthm}
\theoremstyle{plain}

\theoremstyle{definition}

\usepackage{placeins}
\newcommand{\subparagraph}{}
\usepackage{titlesec}
\usepackage{caption}
\theoremstyle{definition}

\theoremstyle{remark}

\titlespacing*{\section}{0pt}{5pt}{5pt}
\titlespacing*{\subsection}{0pt}{3pt}{3pt}
\titlespacing*{\subsubsection}{0pt}{3pt}{3pt}
\begin{document}			
\bstctlcite{IEEEexample:BSTcontrol}
\clearpage
\thispagestyle{empty}
\maketitle

\begin{abstract}
Due to proliferation of energy efficiency measures and availability of the renewable energy resources, traditional energy infrastructure systems (electricity, heat, gas) can no longer be operated in a centralized manner under the assumption that consumer behavior is inflexible, i.e. cannot be adjusted in return for an adequate incentive. To allow for a less centralized operating paradigm, consumer-end perspective and abilities should be integrated in current dispatch practices and accounted for in switching between different energy sources not only at the system but also at the individual consumer level. Since consumers are confined within different built environments, this paper looks into an opportunity to control energy consumption of an aggregation of many residential, commercial and industrial consumers, into an ensemble. This ensemble control becomes a modern demand response contributor to the set of modeling tools for multi-energy infrastructure systems.
\end{abstract}

\section{Introduction}
Energy delivery is a key industrial process that spans across multiple critical infrastructure systems and tightly weaves in many, if not all, residential, commercial and industrial activities. Traditionally, operations of the energy systems – electricity, gas, heat – are separated. Each of these systems operates based on domain-specific assumptions and ad-hoc practices. Interactions between the systems were, and to a large extend still are, minimal, therefore limiting opportunities for inter-system synergies. This paper contributes to the line of research challenging the status-quo. We identify potential for coupling interfaces between the three major energy infrastructures and seek to coordinate their operations in order to improve energy efficiency, reliability and resiliency. Thus far these efforts commonly take an aggregated, system-centric view on operations and therefore often sacrifice details, e.g. spatio-temporal granularity and decision-making hierarchy. As a result of these simplifications, the multi-system couplings are exploited at the very top (system) level and their benefits are acquired and subsequently appropriated only on behalf of the entire system. However, this paradigm runs into a conflict with the current push toward decentralizing infrastructure operations, proliferation of control, monitoring and communication technologies and cost-effective local energy supply means.

Motivated by the proliferation of demand response programs in the electric power distribution sector, this paper seeks multi-system synergistic effects in the context of the hierarchical decision-making structure of the electricity, gas, heat infrastructure systems. Due to the techno-economic similarities between these infrastructure systems, the demand response experience obtained in the electricity context, as well as the modeling methods, can support the development of similar customer engagement frameworks in gas and heat infrastructure systems. In particular, we suggest to delineate the three common levels in each infrastructure – correspondent to utility, aggregator, and consumer, respectively. In turn, this delineation motivated by the real-life practice, allows to select a sufficiently accurate spatio-temporal granularity for each level and to design level-specific couplings. The top/utility level represents infrastructure operations from the system perspective, i.e. as currently adopted by local electricity, gas, and heat utilities, and is customized to account for ubiquitous infrastructure constraints and techno-economic and policy objectives. The middle level is designed to accommodate third-party service providers, e.g. energy retailers or aggregators, that can arbitrage between the utility and customers to leverage emerging information and communication technologies that intend to harvest additional economy-of-scope benefits in addition to economy-of-scale benefits pursued by utilities. The bottom level accounts for the decision-making process of individual customers based on their needs, preferences and often sub-rational choices. However, due to the diversity of customers and their techno-economic features, modeling individual consumers in utility and aggregator operations has proven to be challenging in the current practice. Therefore, it is imperative to create an accurate representation of the aggregated behavior of customers that would adequately capture their individual features and correlations among them. Taken together, these three levels make it possible to thoroughly trace and analyze electricity, gas, and heat flows from local utilities to customers across domains and thus unlock opportunities to seek synergistic operations among and within levels of the critical infrastructure systems involved.

The success of the proposed hierarchical approach hinges on the ability to coherently integrate all three levels in one decision-making framework that can accurately represent each perspective with a sufficient level of operating detail and ensure compatible modeling choices for each level. To this end, the approach builds on the Markov Process (MP) and Markov Decision Process (MDP) theory that has been shown in the past as an efficient framework to operationalize distributed energy resources and energy delivery in a network-constrained environment. The MDP theory is also well-suited to account for model and parameter uncertainty observed by utilities, third-party providers and customers, and to seek consensus decision strategies under different assumptions on communication interfaces between each level. The resulting MDP-based decision-making framework incorporating all three levels will create new and refine existing couplings between interdependent heat, electricity, gas infrastructures and inform on the appropriate spatio-temporal granularity for multi-system aggregation at each level.

This paper makes the following contributions:
\begin{enumerate}
 \item We generalize the use of the MDP for accurately representing dynamics of energy customers at the utility and aggregate level. This representation makes it possible to accommodate the uncertain customer dynamics using stochastic and robust optimization methods. Furthermore, we enhance this MDP framework to learn the optimal control policy for effectively dispatching TCL ensembles using the so-called Z-learning algorithm. This MDP framework is applicable to all energy infrastructure systems.
 \item The proposed MDP formulations are then integrated with the optimal power flow (OPF) problem used by electric power utilities to optimally dispatch available energy resources, while ensuring that all asset and network limits are securely met. The OPF problem accommodates the uncertainty of stochastic generation resources using chance constraints (CC). The integrated problem combining the MDP and CC-OPF optimization is then solved using an iterative algorithm based on the dual decomposition, which allows for co-optimizing the MDP and CC-OPF decisions. While the integrated problem is formulated for the electric power distribution network, it can also be applied for gas and heat networks as noted in the paper. However, we do not present these models due to a lack of real-life data.
 \item The usefulness of the proposed approach is demonstrated on real-life data for both residential and commercial consumers. The presented data-driven use cases are primarily focused on electric power distribution systems but a similar procedure can be extended to the other energy systems if data is available.
 \item Finally, we outline future extensions that can improve and customize the proposed MDP framework for applications in multi-energy systems, including resiliency enhancements, model reductions, and usage beyond electric power distribution networks.
\end{enumerate}

The rest of this paper is organized as follows. Section~\ref{sec:building} describes current practices to model built environments using physics- and data-based approaches. Section~\ref{sec:Multi_energy_inf} summarizes modeling practices for multi-energy infrastructure systems and motivates their enhancement to include built environments from the perspective of utilities, customers, and aggregators using a hierarchical modeling approach. Sections~\ref{sec:markov} and \ref{sec:mdp} describe the proposed hierarchical modeling approach, which uses the Markov Process (MP) to characterize the electricity, gas, and heat consumption of buildings and the MDP to optimally dispatch this flexibility in coordination with infrastructure systems. Section \ref{sec:learning} discus opportunities for learning methods to improve the MDP performance for real-life applications. Section~\ref{sec:future} discusses further extensions of the proposed MDP framework.

\section{Built Environment in Multi-Energy Infrastructure Systems} \label{sec:building}
Previously, multi-energy infrastructure systems have been studied from the perspective of a single or multiple centralized planners or operators, \cite{energy_flow1}-\cite{energy_flow4}. Accordingly, existing modeling and algorithmic solutions focus on a system-centric representation of energy flows, thus only allowing for coordinating respective infrastructure systems at a bulk resource level. This bulk resource level, also known as the energy hub \cite{energy_hub}, aims to store or convert different energy types prior to distributing it to consumers with the main intention of meeting operating limits in each infrastructure and reducing their respective operating costs. Notably, the conversion within a energy hub can be either uni- or multi-directional, which is often implementation specific. To a large extent, previous propositions to jointly operate multi-energy infrastructure have ignored customer-end dynamics, i.e. there was no equivalent of the energy hub within a given built environment. On the contrary, recent advances in the deployment of distributed energy resource and energy efficiency technologies in commercial, industrial, retail, and residential buildings make it possible to internalize the user-end perspective in multi-energy power flow computations, and thus enable a better accounting of edge dynamics in each infrastructure. The opportunity to include edge dynamics in multi-energy flow modeling and operating practices calls for a holistic modeling framework that can represent electricity, gas, and heat consumption within built environment and allow for energy conversion beyond system-level energy hubs.

Although methods to evaluate energy dynamics of built environments buildings are scarce, decentralized energy supply systems having distributed control capabilities are found advantageous over centralized systems to satisfy preferences on end users during normal and contingency operations \cite{amin_2008, goldthau_2014, Panteli_Mancarella_2015}. However, these previous studies view built environment as a whole on an infrastructure level, deeming individual buildings as black boxes, or only focused on individual buildings, ignoring the connection of buildings through infrastructures. However, since cyber-physical interfaces among individual buildings have become more ubiquitous, it is now possible to explore characteristics and energy flexibility of connected buildings (e.g. an ensemble of relatively homogeneous buildings) in urban environments.

However, modeling such ensembles is challenging due to the need to acquire, process, aggregate, and actuate building-specific data. Although one can leverage Building Information Models (BIMs), which are typically available for urban buildings, there is a number of well-recognized and salient challenges. BIMs provide 3D and computer interpretable representation of physical and functional characteristics of building elements (e.g., exterior enclosure, structural columns, material type, geometry, dimensions, connections to other building elements, location, etc.) when queried. Hence, these models have been heavily used in the current practice as well as in research to extract, transform, and build off building data for various practical and research problems. These areas include construction scheduling \cite{goldthau_2014}, cost estimation \cite{lee_2014, nepal_2012, Ergan_Akinci_2012, Staub-French_2003}, design improvement \cite{Jeong_and_Ban_2011}, facility operations \cite{Yang_and_Ergan_2015, Yang_and_Ergan_2016}, and model and code checking \cite{Martins_and_Monteiro_2013}. Although these models have been effective in providing the required data in these application areas, their potential has not been fully explored for multi-energy dispatch. The effectiveness of the models for multi-energy dispatch applications will heavily depend on the data representation and level of detail of the stored data. What data to store, how to represent it and the granularity of storing the data are factors that will change the decision-making procedures at the building and system levels.

Furthermore, assessing the flexibility that each building in the ensemble can provide requires
accounting for electric power, gas and heat dynamics, which are
driven by comfort and behavioral preferences of occupants and exogenous conditions (e.g. temperature, humidity, etc).
Currently, there are two large groups of methods to model and
forecast building electricity, gas and heat consumption: (i) modeling relevant physical processes (e.g. heat transport, electromechanical considerations, Kirchoff’s laws, evaporation, etc)
and (ii) data-driven (e.g. statistical analyses and inference). Physical models not only use available measurements and static
building parameters from BIMs, e.g. location, floor area, number of stories, detailed information on the heating, ventilation, and air
conditioning (HVAC) system, lights, coils, doors and windows,
but also operate with specific models that govern dynamics
of relevant characteristics \cite{four}. EnergyPlus, for example, is a
popular simulation tool for modeling energy needs of buildings
using detailed thermo- and mass- modeling of energy flows
inside the building \cite{five}. EnergyPlus can also be used for an offline and off-site analyses to determine set point adjustments of the energy consumption \cite{six}. The advantage of using the
physics-based models is in their ability to describe buildings
without prior observations. However, the performance of these
models is highly sensitive to the number and accuracy of the
underlying modeling choices and assumptions, as well as to
input parameters. Physics-based models often require more inputs than existing data acquisition systems can provide \cite{four}, and
therefore incur significant uncertainties in both model parameters and dynamic processes. Using such models for controlling
an ensemble of buildings may lead to computational issues that
would prevent their scalability and implementation for real-life
decision-making. Due to these shortcomings, it is common
to sacrifice modeling accuracy of the physics-based models,
which may lead to a loss of their predictive power. On the other hand, in lieu of the physics-based models,
one can use machine learning and statistical modeling to
perform data-driven studies of buildings using a vast amount
of historical data available at the buildings equipped with
smart meters. These models are trained using the historical
energy consumption data and other parameters (e.g. weather
conditions, daily operational schedules, and control functionality) \cite{three, six}. Then, the models can be used continuously
to learn and predict energy usage from previously observed
conditions. Availability of data is crucial for such approaches,
especially when attempting to predict consumption with a
minimum set of required inputs \cite{three}. Notably, such data is
publicly available at an urban scale. For example, New York
City’s Local Law 84 (LL84) requires that all commercial
(including multi-family) buildings of 50,000 square feet or
more must report energy and water consumption on an annual
basis. Although this data is very coarse, it has been used
in a combination with other building information (e.g. year
built, floor area, property-use type, occupancy) to develop
more accurate data-driven building models \cite{seven}. On the other
hand, the data-driven models are data-intensive and building specific and require large amounts of data for re-training or
re-calibration, even when minor changes are made to the
buildings. This hinders scalability of the data-driven models
and their ability to represent an ensemble of buildings with
varying characteristics. Furthermore, numerous studies have
revealed that data-driven models may yield discrepancies (up
to 100\%) between the models outputs and the observed data
\cite{seven, eight}. To reduce the gap between the prediction and the
actual performance, researchers have conducted calibration studies
to tune the various inputs to match the observations \cite{nine, ten, eleven}. Nonetheless, calibration is still an over-specified and
under-determined problem due to a relatively large number
of inputs and a few measurable outputs \cite{three}.

Alternatively, machine learning and data-driven techniques can be leveraged to inform physics-based building models. In this paper, we construct a MP
to represent the energy consumption of building appliances to assess the building flexibility for various applications in
electric power, gas and heat distribution systems. First, the physical
building model is used to characterize the MP using the
probability transition matrix. Based on this MP, we formulate
the MDP that can in turn be
used to optimally control electric and heat appliances within
buildings either by the local utility, third-party aggregators,
or building managers. The ultimate objective of this paper is
to combine the scalability of the MDP with the accuracy of
the physics-based models, and to co-optimize dispatch of the ensemble with multi-energy infrastructure networks.

\section{Multi-Energy Infrastructure}
\label{sec:Multi_energy_inf}
The development of a modeling framework to jointly operate multi-energy infrastructure systems has been investigated over the past decade to pursue a broad range of economic and physical performance goals. To this end, the typical objective is to select steady-state settings of controllable infrastructure assets to minimize the joint operating cost, while accounting for various engineering constraints on infrastructure elements. These modeling frameworks have been used for energy flow, reliability, cross-system optimization, and investment evaluation applications. The energy flow models, which vary in underlying assumptions, modeling accuracy and computational performance, are typically steady state and allow for computing the flows across infrastructure to ensure that energy supply is sufficient to meet the expected energy demand, \cite{energy_flow1, energy_flow2, energy_flow3, energy_flow4}. The reliability applications, \cite{reliability2,reliability1}, are based on energy flow models and account for potential failures of infrastructure elements. Such reliability-motivated models are often probabilistic and account for different likelihoods of failures, repair times and budgets, and risk imposed by each element on the infrastructure systems (e.g., likelihood times impact). Cross-system optimization, \cite{energy_flow2}, utilizes energy flows and reliability models to evaluate how much flexibility (if any) each infrastructure system can provide on different time scale, varying from several minutes to several weeks, and how this flexibility can be efficiently converted between different energy carriers with minimal losses. Finally, the investment evaluation applications use simplified models described above to inform decision makers on economically and technically sound investment decisions to support joint operations of multi-energy systems. Often these models for investment evaluation must account for a number of uncertain externalities and, therefore, render computationally demanding optimization problem.

We consider these applications of multi-energy models for infrastructure systems from the perspective of utilities, customers and aggregators below.

\subsection{Utility's Perspective}

From the utility\footnote{The term utility usually denotes a centralized supply organization that provides a regulated service in a given area. In the US, utilities can be either investor-owned, publicly-owned, cooperatives, or federal. In the following, we use the term utility in a generic notion of the service provider and do not assume a particular ownership structure. The modeling developments presented below are uniformly applicable to all utilities.} perspective, which are likely to be the only real-life entities positioned to operate multi-energy systems in a centralized manner, the main value proposition for adopting multi-energy operating practice stems from improving energy efficiency, reliability and cost savings.

\subsubsection{Electricity supply} 

 The current electricity supply architecture has evolved as the result of the restructuring and deregulation process \cite{Joskow1998} and the introduction of wholesale competition among electricity producers \cite{hogan_2000_wholesale_market, LEWIS20071844}. The current architecture allows for wholesale competition over the transmission network between large generation companies (Genco), whose objective is to sell electricity at the highest price, and distribution utilities, whose objective is to purchase electricity at the lowest price. The utilities supply the purchased or self-produced electricity to consumers using the distribution network. With a few notable exceptions (e.g., CA, NY, TX \cite{Littlechild_retail_analysis}), US electricity consumers are bounded to receive their electricity supply from the utility at a tariff/rate regulated by local authorities.

Both the transmission- and distribution-level electricity supply rely on operating and planning tools based optimal power flow models, which optimize dispatch settings of controllable generation and transmission assets given their operating limits,
forecasted operating conditions, power flow and nodal voltage
constraints, and security margins. In general, OPF refers to a family of decision support tools that seek to optimize a given objective function (e.g. generation cost, total power losses, profit, utility), while ensuring that optimized operations meet the limits imposed by electrical laws for a power network, as well as stability and capacity constraints on bus voltages, generation assets and line flows, \cite{opf_frank}. The recent push toward integrating renewable energy resources with intermittent outputs
has introduced a new degree of uncertainty and complexity in
transmission operations. First, it requires dealing with nonconvex and nonlinear alternating current (AC) power flow equations (based on the power flow constraints given by the Kirchhoff's laws), which make even
the deterministic OPF problem NP-hard \cite{7063278}, i.e. it cannot be solved in polynomial time. Second, it is difficult to model uncertainty propagation throughout the network.
One approach to circumvent those challenges is to replace the
AC power flow equations with the linear direct current (DC) approximation, which neglects power losses, assumes small angle differences,
and parameterizes the voltage magnitudes. The linearity and
convexity of the DC approximation enables the application of
scenario-based \cite{867521}, chance-constrained \cite{doi:10.1137/130910312} and robust \cite{6575173,6948280,6917060}
optimization techniques to deal with the uncertainty of renewable generation resources in a tractable manner but reduces the accuracy of the model. Alternatively, one can use linearized AC power flow approximations, e.g. \cite{deka2017structure,8600344}, or convex (second-order) relaxations, \cite{8481575}, that improve the model accuracy at a modest increase in computing times. Notably, models for distribution power flows can take advantage of typically radial distribution network structures, which allow for linear LindDistFlow and second-order DistFlow formulations, \cite{25627}, that are capable of providing more accurate tractable solutions then DC approximation. Finally, the coordination between transmission and distribution systems has previously
been studied in the operating context with the primary focus on steady state conditions, \cite{TD1, Bragin_2018pes}. Sun et al. \cite{masterslave} formulate a global power flow problem
for the unified transmission and distribution system and solve it
using a master-slave-splitting iterative algorithm. Li et al. \cite{TD2, TD3} propose a decomposition approach for the coordinated economic dispatch of the transmission and distribution systems that
can capture heterogeneous technical characteristics of these systems and reasonably model information flows between them. In
\cite{TD4}, the decomposition algorithm from \cite{TD2,TD3} is improved
to handle ac power flow constraints for both the transmission
and distribution systems. Although such coordination schemes as in \cite{TD4}, make it possible to improve energy efficiency of the power grid, they do not account for the flexibility available at the edge of distribution systems.

\subsubsection{Gas supply} 
To deliver gas one needs to maintain a sufficiently high pressure along the pipe. Pressure is kept at 200--1,500 psi range in the transmission part of the system which is achieved by placing pump stations every 50-70 miles to compensate for the pressure drop. Traditional consumers of gas at this high pressure (transmission) level are city gates. There are also natural gas plants, typically run in co-generation mode, that extract gas from the system. Pipes, pumps, city-gates (local distribution companies) and also gas reservoirs (underground and more modern compressed gas units) form the transmission level network. Network topology at this level contains a very few loops, that is, the transmission network is largely tree-like. On the contrary distribution network, which starts at the city-gates and goes down to house-holds and mid-to-small size businesses, are typically loopy to guarantee resilience (restarting the gas system is expansive, as it requires an expansive manual manipulations by a crew to meet safety standards). Pressure at the distribution level is lower ($0.5-200$ psi), and the reduction from high to lower pressure is achieved in a number of steps at the gate stations. Natural gas system is built to allow significant variations in pressure, with the allowed window often covering up to 50\% of the nominal level. This arrangement allows to run the system in a relatively loose way, i.e. with much less frequent (than in the power system) changes/adjustments. As a result, injection of gas into the system and extraction of gas from the system are not balanced at the time scales of minutes and hours. The balance is restored (in average) at the scale of a day and sometimes multiple days. In this traditional legacy set up, automatic controls are largely local, e.g. seeking to maintain predefined pressure at the pump stations, with periodic manual corrections by the operator. Modern systems are characterised by an increased level of fluctuations in consumption originating from (a) gas-fired power plants which are often used as the first responders on the power system side to mitigate fluctuations caused by renewables, i.e. wind and solar; and from (b) multiple small and medium size active consumers responding not only to external temperature (easy to monitor) but also engaged in an arbitrage of multiple energy resources to meet their heating/cooling needs. Modeling of the gas system operations has been the subject of extensive research over the last 40 years \cite{84Osi,87TT,94Kiu,00ZA,08HMS,15Chertkov,2017Dyachenko}. Optimization and optimal control of the natural gas operations are also discussed extensively in the literature, both in the stationary (planning) \cite{68WL,90LP,00WRBS,10Bor,2012Babonneau,13MFBBCP,2015Zlotnik,2019Sukharev} and more recently in the operational (dynamic, accounting for line packing) contexts \cite{2016Mak,2017Zlotnik}. However, these studies remain largely academic, i.e. not yet implemented in practical operation and planning of the natural gas systems.
\subsubsection{Heat supply} 
District Heating Systems (DHS) are built to resolve heating needs of many geographically collocated residential consumers in a centralized way. DHSs are wide-spread in European and some American cities in the northern hemisphere that experience significant seasonal variations. While the first DHSs, built in NYC, Chicago, Seattle and Paris, operated on steam, DHSs of the third generation, largely adopted in Nordic and other European countries, are much more efficient (run under 70 degrees of Celsius), use plastic pipes (no corrosion), and are operated at slower velocities (thus leading to longer delays in the heat delivery from the sources to consumers). Even though modern DHSs utilize automatic controls at the heat sources, pumps and some consumers, system-wide adjustments are set in action by human operators. The controls are either hydraulic, changing mass flow, or thermal, achieved through heating/cooling at the sources. When compared with other energy infrastructures, DHSs show much stronger dependence on external conditions (outside temperature, wind, cloud coverage) and thus show more significant variations in the operational conditions. Slower flows and active response of consumers are other contributors to the growth of DHS variability as it becomes modern. In terms of the temporal scale of operation, DHSs are roughly as inertial as the Natural Gas systems, thus they provide a significant demand response balancing potential to the power system operated at the same district/distribution level. (See \cite{2019NNN} for further details on the network modeling, parameter identification, control and optimization aspects of the DHS technology.)


\subsection{Consumer's Perspective}
Motivated by the roll-out of smart grid technologies, which includes communication and control means allowing for alternating standard consumption patterns, consumers can be incentivized to consume energy of different types in a way that alleviates bottlenecks in operating infrastructure systems. In return, consumers can expect a certain re-numeration that would compensate for any discomfort incurred by reducing, shifting, or eliminating their consumption. However, several challenges exist that limit participation levels of customers in demand response program, \cite{OCONNELL2014686}. First, massive enrollment of demand-side participants in such programs is only possible if their premises are equipped with proper metering and automatic control units that can seamlessly communicate with the utility and on-demand execution of desired commands. Although these units are commercially available, their upfront cost is still prohibitively expensive and their wide deployment is limited to large metropolitan areas. For instance, Consolidated Edison (ConEd) - a distribution company in New York has began installation of smart meters in 2017 in Staten Island, which is expected to continue through 2022 until it covers Brooklyn, Manhattan, the Bronx, and Queens. However, even if the smart metering and automatic control systems are available, there is a number of issues that may lead to low participation rates. First, among potential demand-side participation, the perception is that the value proposition of such programs is relatively small as compared to electricity tariffs, self-valuation of comfort and rent (mortgage) rates. Second, customers typically have low awareness of their potential to improve energy efficiency and associated benefits to the environment and society as a whole.

Extending these customer-end challenges to the multi-energy context renders it difficult to seamlessly exchange energy of different types at the customer level, which motivates the use of aggregation techniques and control methods described below.


\subsection{Aggregator's Perspective}
Due to the complexity of aggregating small-scale generation and demand-side resources, power utilities may not fully harvest benefits of these resources since their business models and practices seek the economy of scale benefits. As a result, there is an opportunity for demand-side aggregators to act as a mediator between the utility and consumers with demand-side and generation resources, thus collecting the benefits of the economy of scale (e.g., by reducing transaction costs, providing better information and other services that are prohibitively expensive for individual Distributed Energy Resources (DERs)) and economy of scope (e.g., via providing multiple services to the utility), \cite{BURGER2017395}. From the utility perspective, the advantage of engaging in interactions with aggregators is in replacing the need to interface and support continuous communication with each demand-side and generation resource. Instead, the utility has to deal with a rather small number of aggregators that in turn coordinate their DER portfolios based on the utility's instructions and accumulate risks on the performance of individual consumers. Thus, in addition to streamlining communication, aggregators can hedge financial risks for both utilities and DERs, as well as multiple competing aggregators may reduce volatility in electricity prices \cite{BURGER2017395}. However, even though competitive forces should theoretically encourage aggregators to deliver cost-effective services to DERs and utilities, there is a concern that existing imperfections (e.g., metering systems or reliability requirements \cite{Borenstein_2015,BURGER2017395}) will reduce, if not completely eliminate, the benefits of the aggregators. Hence, it is important to equip aggregators with decision-making tools that are capable of accurately representing electricity, gas and heat dynamics in built environments.

\begin{figure}[!t]
\centering
\subfloat[]{\includegraphics[trim={0.2cm 1.5cm 0.2cm 0},height=5.0cm,width=0.95\columnwidth]{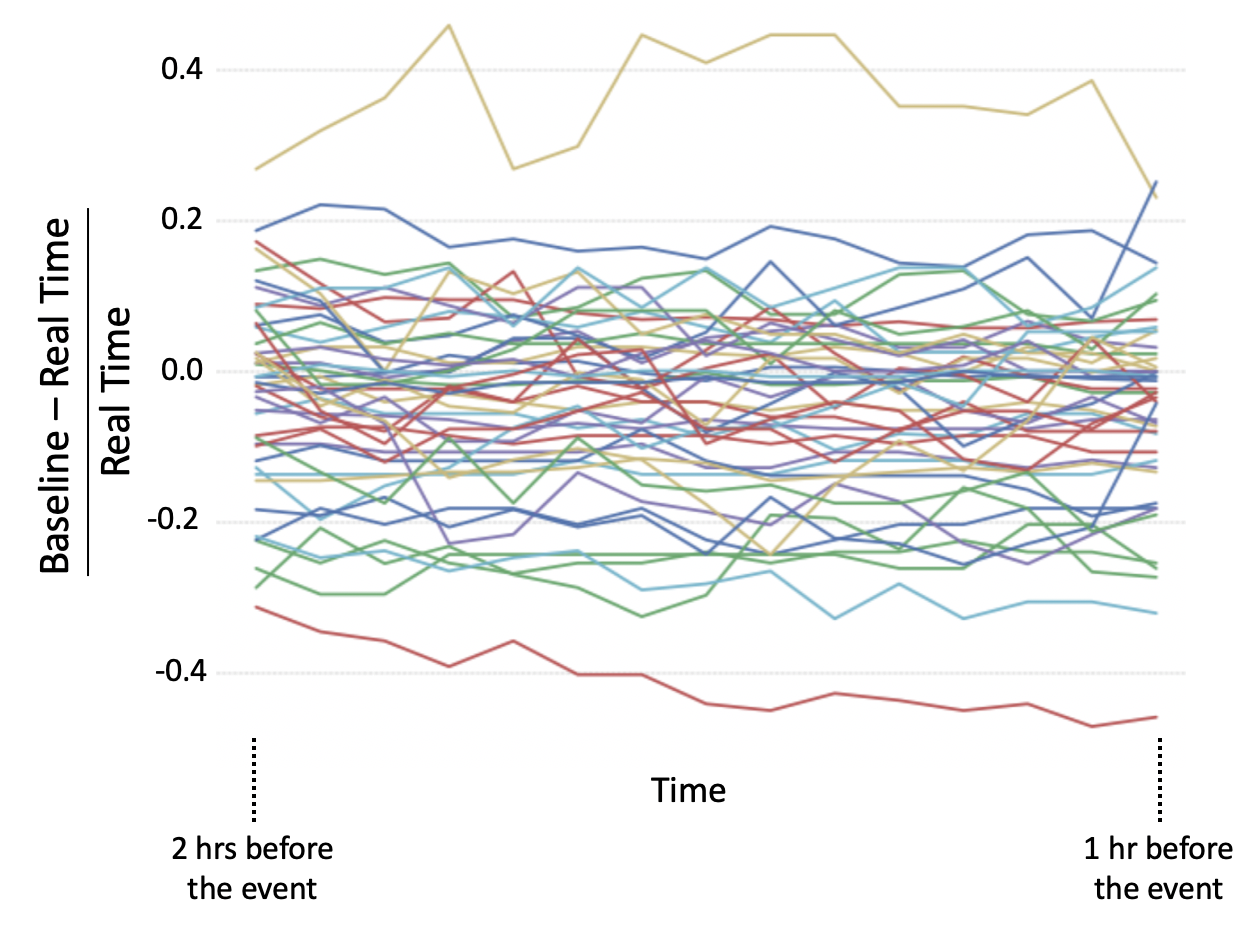}}\\
\subfloat[]{\includegraphics[trim={0.2cm 1.5cm 0.2cm 0.5cm},height=5.0cm,width=0.95\columnwidth]{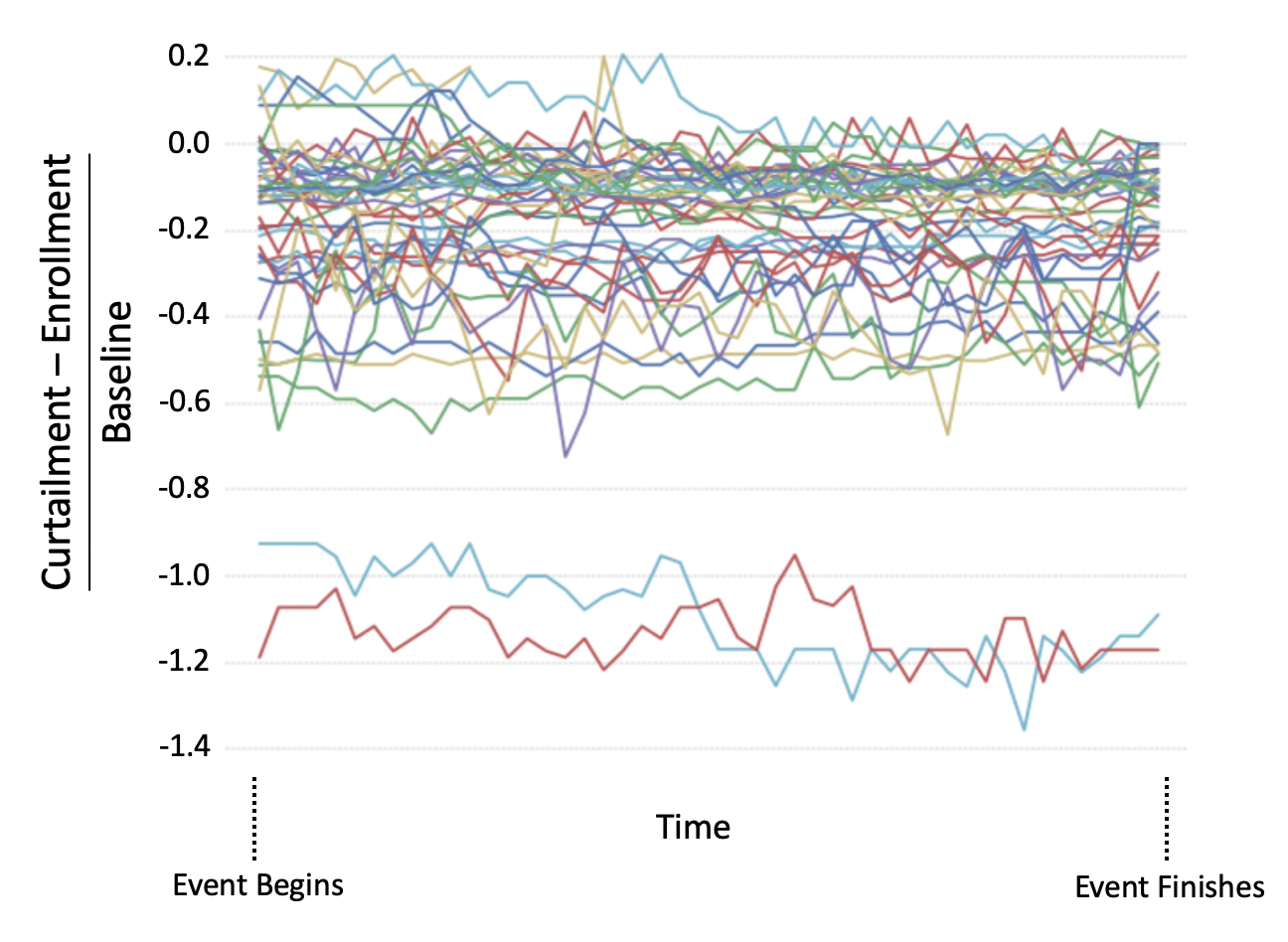}}
\caption{Relative error of (a) baseline against real time; and (b) enrollment against baseline during ConEd DR events. The baseline for ConEd is calculated by averaging the usage of each hourly interval of the top 5 days out of last 10 eligible weekdays. Each line denotes a building before/during the DR events with colors differentiating buildings and events (12 buildings during historical DR events from 2016 to 2018)}.
\label{fig:error}
\end{figure}

Current aggregation techniques use overly conservative methods to estimate the flexibility that can be extracted from demand-side participants. Typically, this flexibility is estimated with respect to its electricity \textit{baseline}, i.e. expected electricity consumption before the Demand Response (DR) event, and \textit{curtailment}, i.e. the expected reduction in electricity consumption during the DR event. These aggregation techniques however neglect the complex dynamics among correlated electricity, heat and gas consumption within a given built environment, which leads to a great discrepancy between the expected and actual performance of DR programs. Fig. \ref{fig:error} compares the accuracy of baseline and curtailment estimations for the DR program operated by ConEd of New York on 12 buildings of the New York University (NYU) campus, where baseline errors can be as great as $\pm$40\% and the curtailment errors vary from -140\% to 20\%. Notably, the baseline error distribution in Fig.~\ref{fig:error}(a) is nearly zero-mean and symmetric, while the curtailment error distribution in Fig.~\ref{fig:error}(b) is skewed toward negative errors, i.e. the enrollment is over-estimated and less capacity is delivered in real-time during the DR event than anticipated.

\subsection{Hierarchical approach} \label{sec:hierarchy}
To improve energy efficiency of multi-energy infrastructure systems, this paper proposes a hierarchical approach that makes it possible to aggregate the energy flexibility of electric, heating, and gas appliances at the customer level and coordinate their usage with infrastructure operations. In practice, however, aggregation programs have become particularly wide spread in electric power systems and, to a large extent, remain in their infancy in other energy infrastructure systems. Nevertheless, due to techno-economic similarities among electric, gas and heat systems, such aggregation techniques will become relevant as more gas and heat customers engage in bidirectional interactions with their gas and heat utilities. Fig.~\ref{fig:DR_communication} illustrates the proposed hierarchical approach for three actors (utility, aggregator, demand response participant) and two decision-making levels. The top decision-making level includes either the utility or aggregator decision maker that needs to estimate the amount of flexibility that can be extracted from demand-side resources. The low decision-making level includes an array of demand-side participants that occupy various built environments and are engaged in two-interactions with either the utility or aggregator at the top level.

The hierarchical approach in Fig.~\ref{fig:DR_communication} is motivated by two key factors. First, it is supported by the available and foreseeable communication infrastructure that can be used toward aggregating distributed energy and demand resources at scale. Second, it fits the regulatory framework currently dominating in the majority of US states, where regulated tariffs with a coarse spatio-temporal resolution (e.g., time-of-use tariffs with peak and off-peak steps) make it possible for profit-seeking aggregators to arbitrage between fine-tuned knowledge of customers' self-valuation of their energy consumption and regulated electricity, gas, and heat tariffs in the system.

\subsubsection{Communication Infrastructure for Aggregation} 
\begin{figure}[!t]
 \centering
 \includegraphics[width=1\columnwidth, clip=true, trim= 0mm 0mm 0mm 0mm]{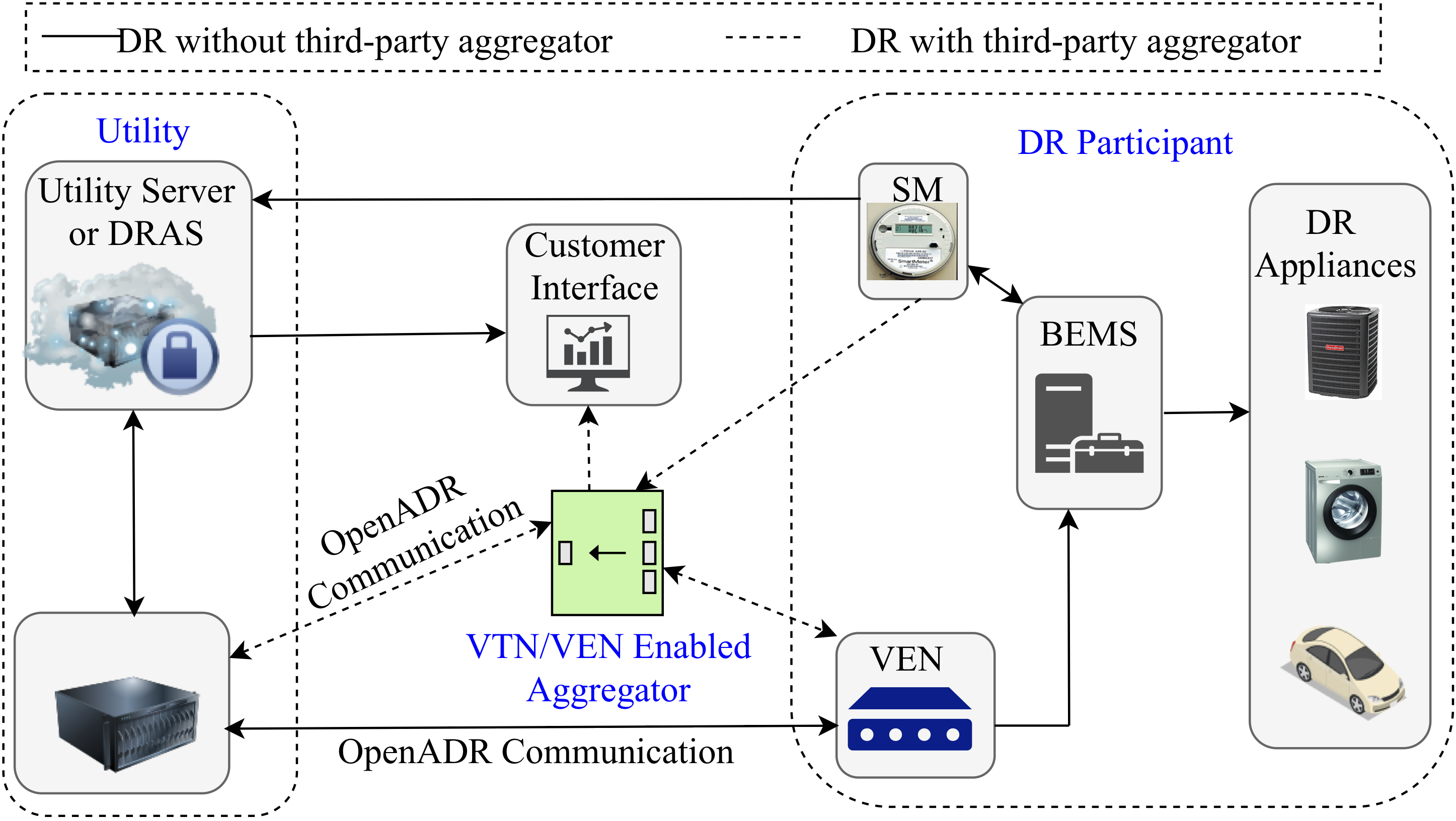}
 \caption{Interaction among utility, aggregator, and DR participants in DR framework.}
 \label{fig:DR_communication}
\end{figure}

\begin{figure}[!t]
 \centering
 \includegraphics[width=1\columnwidth, clip=true, trim= 0mm 0mm 28mm 0mm]{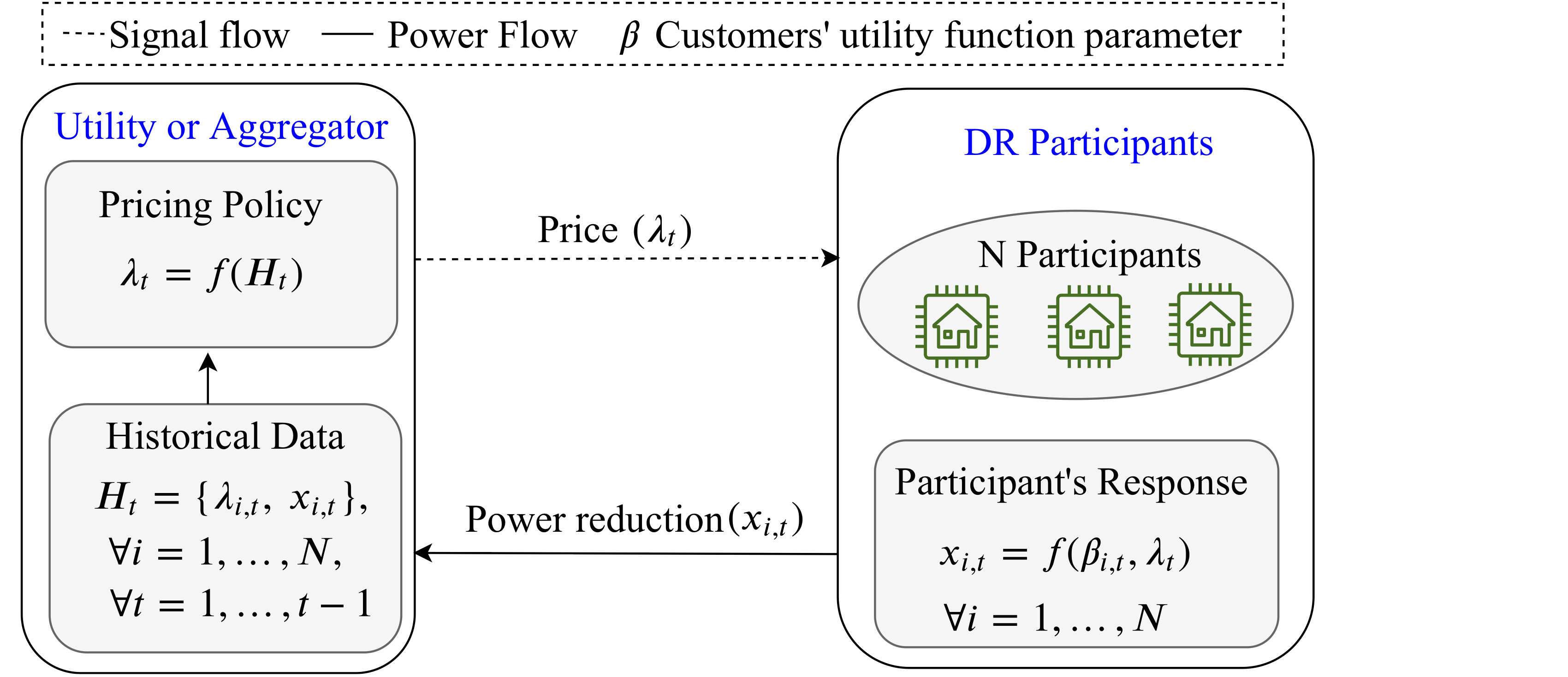}
 \caption{Exploration of DR price in an online fashion using historical data on DR participants' response $(x~kW)$ and DR price $(\lambda~\$/kWh)$.}
 \label{fig:OL_framework}
\end{figure}

The schematic diagram for the Automated Demand Response (ADR) is shown in Fig.~\ref{fig:DR_communication}. The utility acquires DR participant's real-time energy usage from Smart Meters (SMs) installed at the end of DR participants. This energy usage data acquaintance is done via Wide Area Network (WAN). The utility server or Demand Response Automated Server (DRAS) executes DR scheduling and DR pricing algorithm (e.g. online learning based pricing for DR, see Fig. \ref{fig:OL_framework}) using the data collected from the smart meters. The DR pricing signals and schedules are sent to the DR participants using OpenADR 2.0, a non-proprietary, open standardized information exchange model for DR. This model has been recently recognized as an IEC standard 62746-10-1 for the interface between the DR participants and the utility.

The OpenADR communication protocol has Virtual Top Node (VTN), which transmits the DR schedules and pricing signals to Virtual End Node (VEN) and receives response to the DR event from VENs. The VEN coordinates with the local Building Energy Management System (BEMS) and automatically control the high-wattage power appliances such as Thermostatically Controlled Loads (TCLs), plug-in electric vehicles, and washing machines that are registered in DR programs. The interaction among the utility and the DR participants can be with or without the third-party aggregators in between. In the ADR framework with the aggregators, the aggregators combine the DR resources from various customers and use proprietary communication protocols to communicate with the customers' VEN and SMs. However, the aggregators communicate with the utility using OpenADR 2.0 communication protocol. The aggregators provide the DR services to the utility for sharing certain percentage of profit generated by the utility from the DR program. In both the schemes of the ADR (with or without aggregators), customers can get notifications about DR events, track their participation in DR and possible improvements using customer interfaces such as proprietary web-sites and smartphone applications.

\subsubsection{Real-life Aggregation Programs}\label{sec:SmartAC} There is an increasing number of utilities and third-party aggregators enabling DR programs. For instance, ConEd has launched the SmartAC program in New York City (NYC), NY. The program allows its customers to participate in DR programs and reduce the power consumption of their Air-Conditioners (ACs) during peak hours in summer (typically from May through September). Currently, residential and small business customers are enrolled in the SmartAC program. With over 7 million ACs in the NYC area, this program is anticipated to expand substantially shortly. ConEd enables existing window AC to participate in the SmartAC program by installing a home WiFi connected SmartAC kit in the AC. Furthermore, the program accepts various WiFi-enabled ACs that have an in-built SmartAC kit. The kit establishes communication among ConEd, DR participants, BEMS, and AC unit, which is of a similar architecture as shown in Fig.~\ref{fig:DR_communication} (SmartAC kit as VEN installed in the AC). The SmartAC program has a dedicated smartphone application that interfaces ConEd and the DR participants. ConEd sends signals on DR event time, duration, and incentives to the participants via the smartphone application. In return, the participants can respond to ConEd conforming to their participation and can track their DR performance. Furthermore, the participants can control their AC remotely via the smartphone application, even when the DR events are not called. Receiving the consent of the participants, ConEd increases the operating temperature of the AC unit or turns it OFF. Based on the participation in the DR events, the participants earn incentives in the form of \textit{cool points}, which can be redeemed (1000 \textit{cool points} = \$1). Currently, the SmartAC program has static approach to incentives, unlike the one presented in Fig.~\ref{fig:OL_framework}. Customers get flat rebates during the DR enrollment and participation in the called DR events. For instance, customers will get \$10 per AC unit if they allow installing a free SmartAC kit in an existing window AC, \$11 per AC unit on re-enroll, and \$100 per AC unit on enrolling with WiFi-enabled AC. During the DR events, a participant gets \$2.5 per hour per AC unit, while gets \$5 per hour per AC unit if participates for all DR events called by ConEd. There are more advanced DR programs than that ConEd currently practices. For example, ADR programs in California also allow DR via third-party aggregators besides direct interaction with individual DR participants. The aggregators manage multiple DR participants and interact with both DR participants and utilities, and hence, provide competitive DR in scale.

Note that providing solely economic incentives might not be sufficient to persuade customers with bounded rationality, \cite{8810419}, to change their consumption behaviour. In addition to the economic incentives, the utilities can persuade customers with increasing awareness regarding environmental and climate change mitigation impacts to convince more customers to enroll in energy saving programs. For example, the case study carried out among the University of California, Los Angeles (UCLA) students \cite{env_energy_cons} reveals that the group of students who received messages about their energy consumption paired with negative environmental and health impacts consumed less electricity (by 8.9\%) than the group of students who only received messages about their electricity consumption and its cost. Other forms of non-monetary incentives can include magazine subscription, movie and lottery tickets which are offered to customers in return to reducing their peak consumption, \cite{energy_coupons}. In the future, as more energy customers become flexible, considering social values will play a more important role in accurately aggregating TCL customers. Finally, incentive programs similar to \cite{energy_coupons,env_energy_cons} can also be extended to other network energy and transportation infrastructure systems (e.g. for road traffic management, \cite{transit_Balaji,traffic_Balaji}).

\subsubsection{Learning Arbitrage Opportunities} An online learning framework of the DR is presented in Fig. \ref{fig:OL_framework}. The utility or an aggregator sends a pricing signal $\lambda_t$ for a DR event scheduled at time $t$. Upon receiving $\lambda_t$ ( $\forall i=1,\ldots, N$, where $N$ is the total number of DR participants), the DR participant modifies its electricity usage by $x_{i,t}$ kW, either automatically using a pre-programmed algorithm in BEMS or manually via human intervention. It is noteworthy that a DR participant enrolled in DR program can opt out from providing the response, however incurs a penalty. Because of idiosyncratic behavior of DR participants, the electricity demand reduced by a DR participant $i$ during a DR event at time $t$, $x_{i,t}$, is a linear function of $\lambda_t$, with DR participant specific coefficients given by \cite{khezeli2017risk, mieth2018online, li2017distributed}:
\begin{equation}
\label{eq:x_DR}
 x_{i,t}(\lambda_t)=\beta_{1,i}\lambda_t +\beta_{0,i},
\end{equation}
where $\beta_{1,i}$ and $\beta_{0,i}$ refers to the parameters of the cost function of a DR participant $i$. These parameters are unknown to a utility or an aggregator. Hence, the utility learns these parameters using historical data ($H$) on price signals and DR provided, i.e., $H_t =\{\lambda_{\tau},~x_{i,\tau}\}, ~ \forall \tau=1,\ldots t-1, ~ i=1,\ldots, N$. Then the utility explores an optimal $\lambda_t$ for the DR event at current time $t$ and broadcasts to the DR participants.

Successfully implementing demand response programs in practice requires modeling solutions that support resource aggregation with low communication overheads and options for performance improvements via learning. In the rest of this paper, we show that the MP and MDP frameworks are well positioned to meet these requirements.

\section{Markov Process and Data}
\label{sec:markov} 
\begin{figure}
\centering
 \begin{tikzpicture}[font=\footnotesize]
 \node[state,fill=gray!20!white,minimum size=0.5cm] (s1) {1};
 \node[state,fill=gray!20!white,
 left=1cm of s1,minimum size=0.5cm] (s2) {2};
 \node[state,fill=gray!20!white,
 left=1cm of s2,minimum size=0.5cm] (s3) {3};
 \node[state,fill=gray!20!white,
 left=1cm of s3,minimum size=0.5cm] (s4) {4};
 \node[state,fill=gray!20!white,
 below=2cm of s4,minimum size=0.5cm] (s5) {5};
 \node[state,fill=gray!20!white,
 right=1cm of s5,minimum size=0.5cm] (s6) {6};
 \node[state,fill=gray!20!white,
 right=1cm of s6,minimum size=0.5cm] (s7) {7};
 \node[state,fill=gray!20!white,
 right=1cm of s7,minimum size=0.5cm] (s8) {8};
 \coordinate[below of=s5, yshift=0.15cm] (c1);
 \coordinate[below of=s8, yshift=0.15cm] (c2);
 \draw[>=latex,every loop,fill=black!70,
 draw=black!70,
 auto=right,
 line width=0.5mm]
 (s1) edge[line width=1mm] (s2)
 (s2) edge[line width=1mm] (s3)
 (s3) edge[line width=1mm] (s4)
 (s4) edge[line width=1mm] (s5)
 (s5) edge[line width=1mm] (s6)
 (s6) edge[line width=1mm] (s7)
 (s7) edge[line width=1mm] (s8)
 (s8) edge[line width=1mm] (s1)
 (s1) edge[loop above,line width=0.7mm] (s1)
 (s1) edge[bend right, auto=right] (s3)
 (s1) edge[bend right, auto=right,looseness=1.3,line width=0.1mm] (s4)
 (s1) edge[line width=0.1mm] (s5)
 (s1) edge[line width=0.3mm] (s6)
 (s1) edge[line width=0.5mm] (s7)
 (s1) edge[bend left, auto=left,line width=0.1mm] (s8);
 \draw [-,thick,dashed, line width=0.25mm] (s5) -- (c1) node [near start] {};
 \draw [-,thick,dashed, line width=0.25mm] (s8) -- (c2) node [near start] {};
 \draw [{<[scale=1.1]}-{>[scale=1.1]},thick, line width=0.25mm,anchor=south] (c1) -- (c2) node [midway] {Dispatch range};
 \end{tikzpicture}
 \vspace{4pt}
 \caption{MP representation of the TCL ensemble with eight discrete states displaying all possible transitions from state $1$, \cite{Hassan_TCL}.}
 \label{MDP_discretization}
\end{figure}
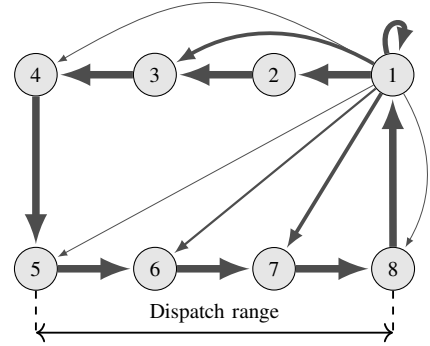

We represent the built environment at the edge of energy infrastructure systems via an ensemble of TCLs. In turn, using the TCL ensemble makes it possible to leverage a MP to represent the electricity, gas, and heat consumption via a given number of discrete states, where each state has an associated energy level. This MP is considered over a finite time horizon with discrete time periods such that the duration of the horizon and time periods co-align with typical time scales of infrastructure operations (week-, day-, hour-ahead). The states are obtained by discretizing the given operating range, which vary for each ensemble based on the operating characteristics of TCLs included, and can be done either uniformly or non-uniformly. Note that such a MP can be constructed for electricity, gas and heat consumption individually or combined, if one accounts for interdependence between electricity, heat, and gas consumption patterns in each built environment. Fig.~\ref{MDP_discretization} illustrates a MP with 8 states with possible transitions from state 1 to other states. Note that the ensemble can remain in the same state.

\subsection{Construction of the Markov Process} \label{sec:mp_construction}
\begin{figure}[!t]
\centering
\includegraphics[width=\columnwidth]{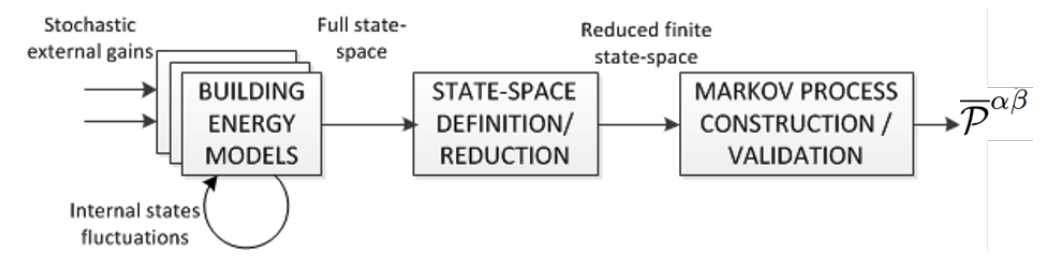}
\caption{The proposed three-step procedure to construct MP for optimal ensemble control of buildings, \cite{Roman_MDP}.}
\label{fig:mdp_three_steps}
\end{figure}

Given the physically accurate building energy and information models discussed in
Section~\ref{sec:building}, we describe a procedure to construct the MP that can be used to formulate and solve the MDP in \cite[Eqs.~(1)-(5)]{Roman_MDP}.
The procedure is illustrated in Fig. \ref{fig:mdp_three_steps} and includes the following
three steps: (i) building data generation and aggregation, (ii) state-space definition and reduction, and (iii) construction and validation of the resulting MP. These steps are further detailed below.

\begin{enumerate}
 \item Building data generation and aggregation: The ensemble is developed by aggregating homogeneous buildings in close proximity, where each building is subjected to some external gains that are stochastic in nature such as outside temperature and wind. The physical building model in \cite[Eqs.(6)-(14)]{Roman_MDP} is applied to the set of buildings to develop building energy models.
 \item State-space definition and reduction: After developing building energy models, a multi-dimensional state-space model is developed for the buildings where state represents dynamics of the building. The state-space is then reduced to a low dimensional model based on those parameters that have a meaningful physical process between them and are controllable such as indoor temperature and power consumption. The parameters in the reduced state-space model are also mutually dependent and important to the consumers in the form of comfort (indoor temperature) and electricity bill (power consumption). Given a reduced state-space model, the authors choose power consumption as a parameter to deal with and translate other parameters into power since building provides services to the grid by shifting its power consumption pattern. The state is then defined in terms of power and discretized into numerically ordered ranges to represent its dispatch ranges and move towards a discrete space MDP.
 \item Construction of the MP: Based on the reduced state-space model and their dispatch ranges, MP is constructed describing the steady state evolution of the system. The normal transition probabilities are computed by tracing and normalizing all the transitions between different discretized states.
\end{enumerate}

\subsection{Application to Residential Households}
\begin{figure}[!t]
\centering
\subfloat[]{\includegraphics[height=3cm,width=0.4\columnwidth]{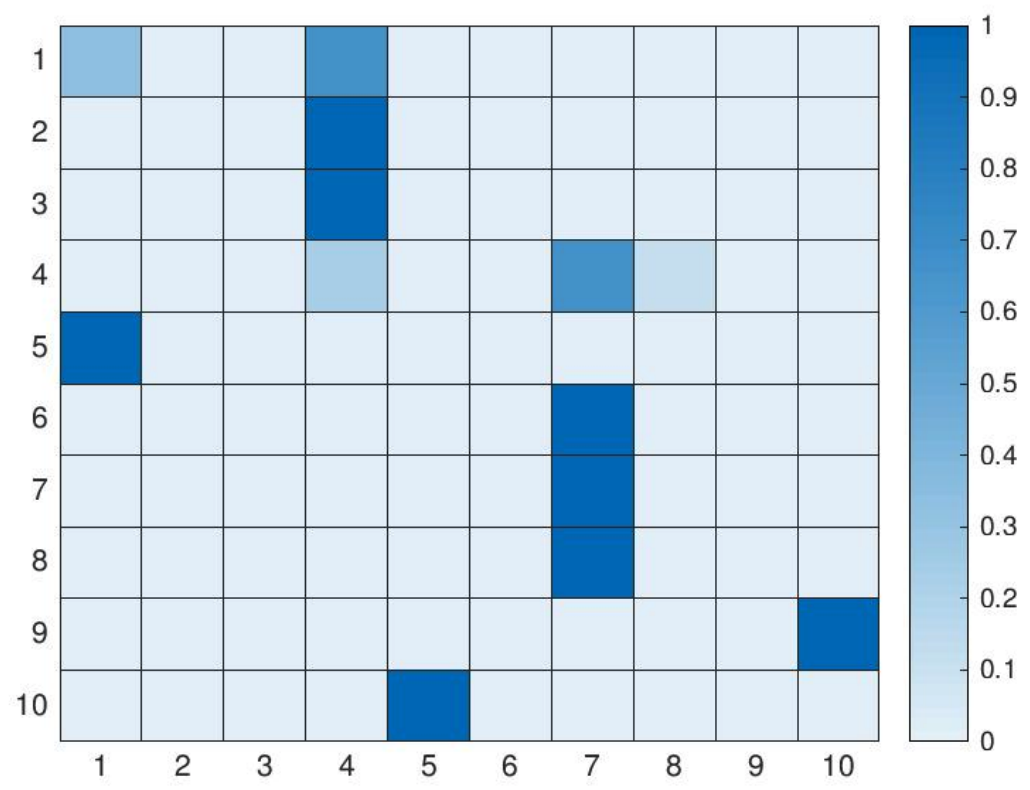}\qquad}
\subfloat[]{\includegraphics[height=3cm,width=0.4\columnwidth]{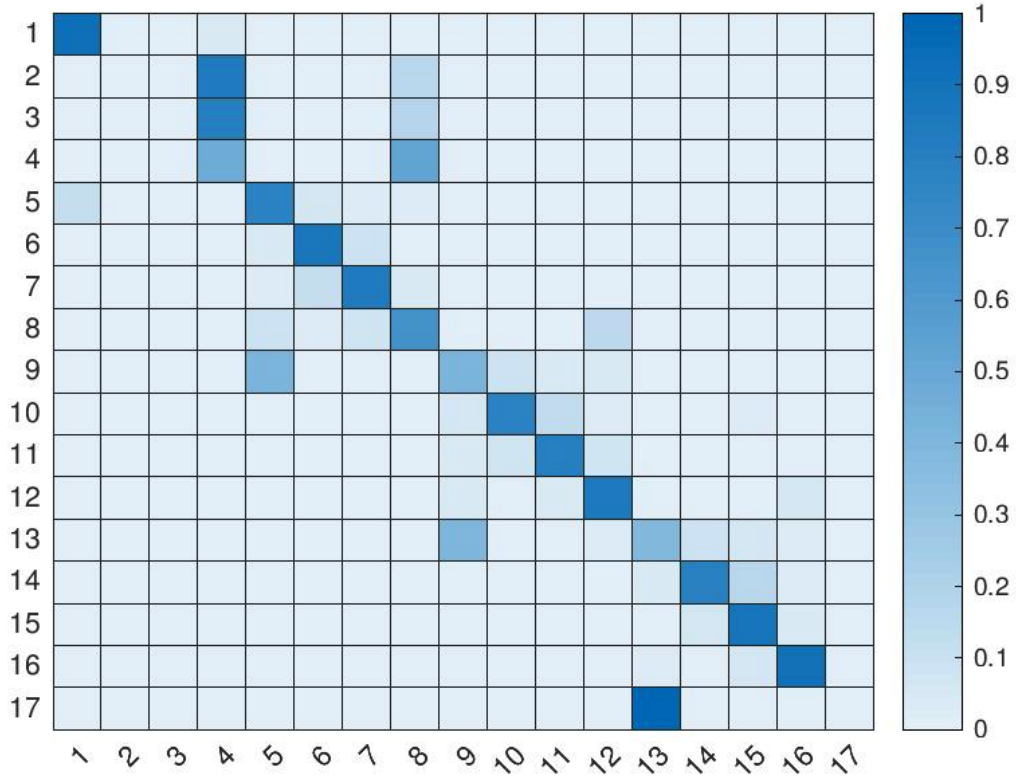}}
\caption{Transition probability matrix for Case 1 with (a) 10 states, and (b) 17 states.}
\label{fig:mp_roman_case1}
\end{figure}

\begin{figure}[!t]
\centering
\subfloat[]{\includegraphics[height=3cm,width=0.4\columnwidth]{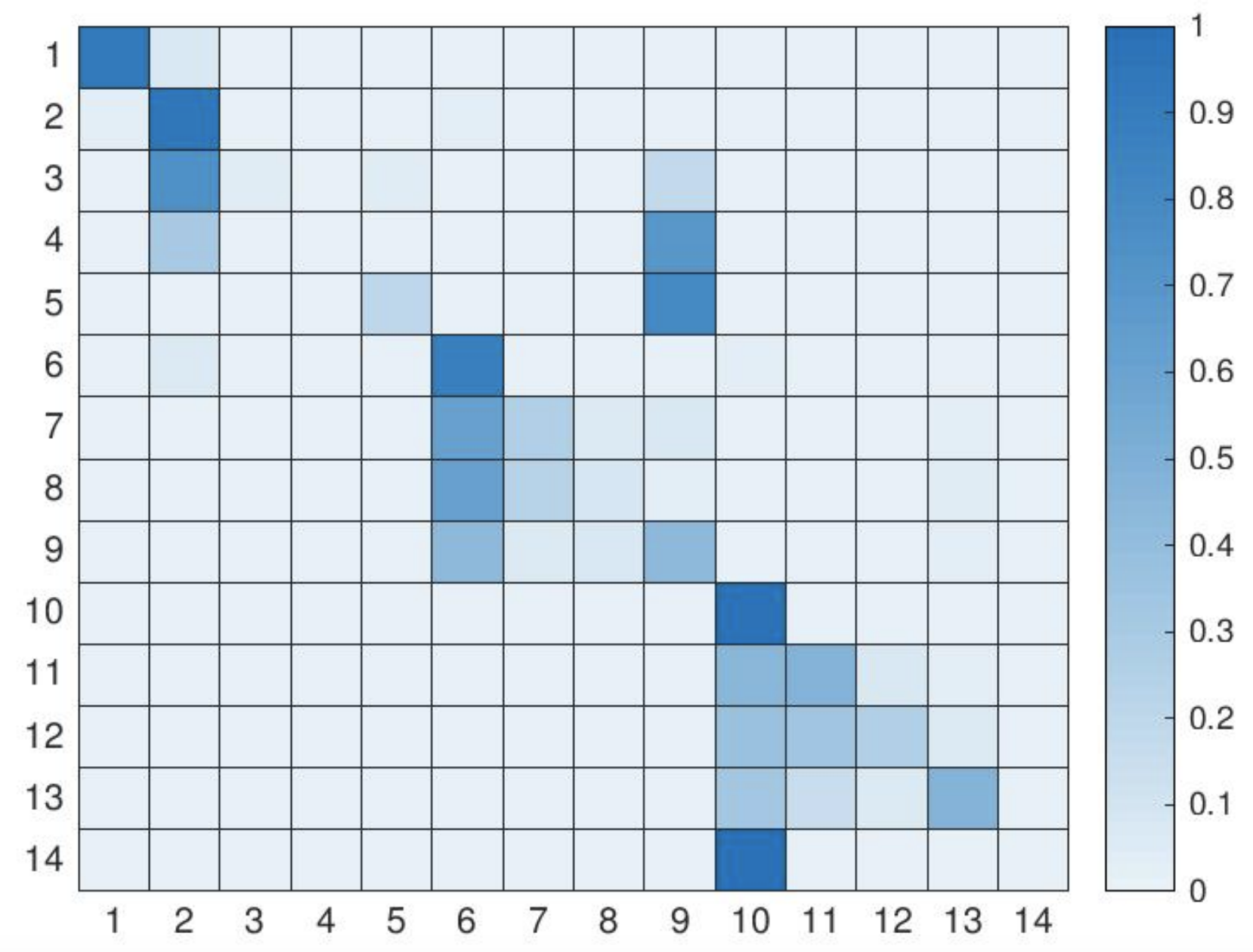}\qquad}
\subfloat[]{\includegraphics[height=3cm,width=0.4\columnwidth]{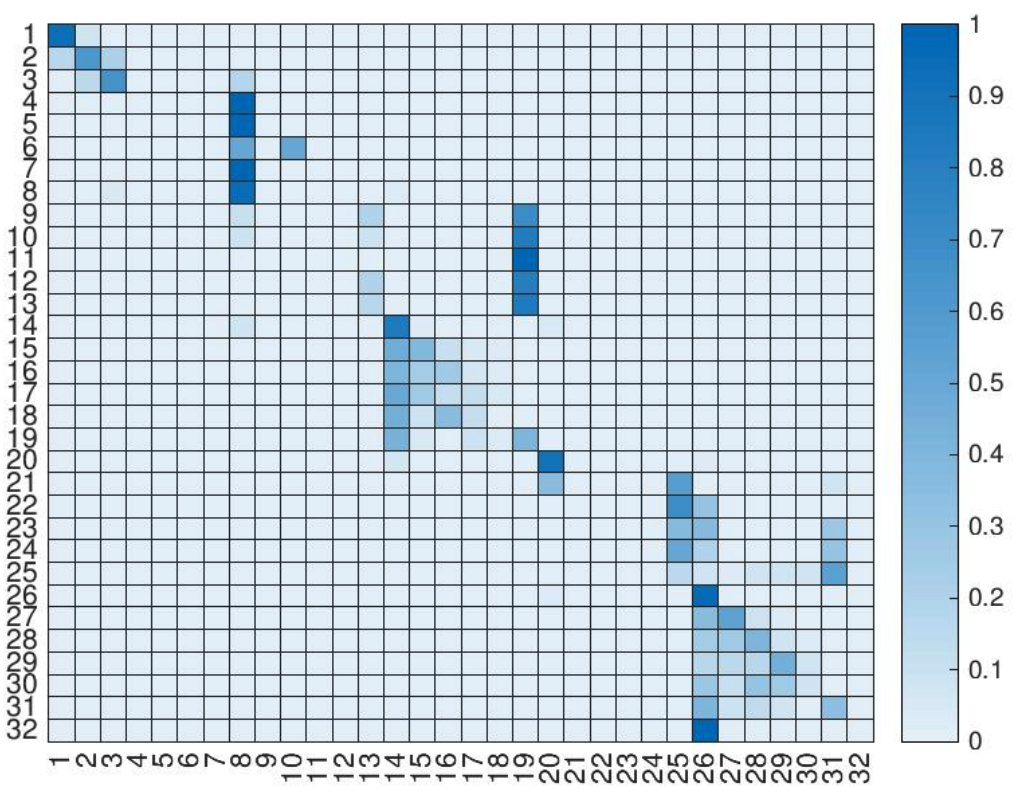}}
\caption{Transition probability matrix for Case 2 with (a) 14 states, and (b) 32 states.}
\label{fig:mp_roman_case2}
\end{figure}

Using the three-step procedure described in Section~\ref{sec:markov}-\ref{sec:mp_construction} and illustrated in Fig.~\ref{fig:mdp_three_steps}, we construct the MP for a portfolio of residential households in Belgium, where each household is an ‘average’ low-energy building, in which the
day and night zones have a surface area of 132 $m^2$
and 138 $m^2$, respectively, \cite{Roman_MDP}. In this built environment, individual heating systems consist of an air-coupled heat pump and a back-up electric resistance heater, which
supply heat to the floor heating system in the day and night
zones, i.e. space heating, and to the storage tank for
domestic hot water. Since different operating principles and low-level controls are available for the heat pumps and auxiliary heaters, we consider two cases specific for residential households.
In Case 1, MP represents the power consumption of the heat pump and ignores the auxiliary heater, while in Case 2, MP represents the power consumption of the auxiliary heater and ignores the heat pump. Based on our experiments calibrating the accuracy of the resulting MP relative to the original physical model, we find that either 10 or 17 states are required to represent the MP in Case 1 (see Fig.~\ref{fig:mp_roman_case1}). Note that considering more states in the MP leads to a sparse matrix, which inhibits further computations. Similarly, the MP in Case 2 requires either 14 or 32 states as shown in Fig.~\ref{fig:mp_roman_case2}. Comparing Figs.~\ref{fig:mp_roman_case1} and \ref{fig:mp_roman_case2} shows the difference in the consumption patterns of heat pumps and auxiliary heaters.

\subsection{Application to a Commercial NYU building}

\newcolumntype{M}[1]{>{\centering\arraybackslash}m{#1}}
\begin{table*}[t]
 \centering
 \caption{Comparison of MDP solutions for different methods}
 \label{table_comparison}
 \begin{tabular}{M{1.5cm} M{2.5cm} M{4.4cm} M{7.5cm}}
 \toprule[.1em]
 \textbf{Method} &\textbf{Uncertainty on Transition Probability} & \textbf{Policy} &\textbf{Value Function} \\
 \midrule[.1em]
 Standard MDP & None & $\mathcal{P}_{t}^{\alpha \beta} = \frac{\overline{\mathcal{P}}^{\alpha \beta}z_{t+1}^{\alpha}}{\sum_{\alpha \in \mathcal{A}}\overline{\mathcal{P}}^{\alpha \beta}z_{t+1}^{\alpha}}$ &$\varphi_{t+1}^{\alpha}$=$-U_{t+1}^{\alpha}-\gamma\text{log}\big( \sum_{\nu\in\mathcal{A}}\text{exp}\big( \frac{-\varphi_{t+2}^{\nu}}{\gamma}\big)\overline{\mathcal{P}}^{\nu \alpha} \big)$\\
 \addlinespace
 Stochastic Extension & Normally distributed &$\mathcal{P}_{t}^{\alpha \beta} = \frac{\overline{\mathcal{P}}^{\alpha \beta}z_{t+1}^{\alpha}\text{exp}\big(\frac{-\sigma^2}{2({\overline{\mathcal{P}}^{\alpha\beta}})^2}\big)}{\sum_{\alpha}\overline{\mathcal{P}}^{\alpha \beta}z_{t+1}^{\alpha}\text{exp}\big(\frac{-\sigma^2}{2({\overline{\mathcal{P}}^{\alpha\beta}})^2}\big)}$ &\!\!$\varphi_{t+1}^{\alpha}$=$-U_{t+1}^{\alpha}\!-\gamma\text{log}\big(\! \sum_{\nu\in\mathcal{A}}\text{exp}\big( \frac{-\varphi_{t+2}^{\nu}}{\gamma}\big)\overline{\mathcal{P}}^{\nu \alpha}$ $\text{exp}\big( \frac{-\sigma^2}{2({\overline{\mathcal{P}}^{\nu\alpha}})^2} \big)\! \big)$ \\
 \addlinespace
 Robust Extension & Normally distributed with ambiguous parameters &$ \mathcal{P}_{t}^{\alpha \beta} = \frac{\underline{\Gamma}z_{t+1}^{\alpha}\text{exp}\big(\frac{-\overline{\hat{\zeta}}}{2(\underline{\Gamma})^2}\big)}{\sum_{\alpha}\underline{\Gamma}z_{t+1}^{\alpha}\text{exp}\big(\frac{-\overline{\hat{\zeta}}}{2(\underline{\Gamma})^2}\big)}$ &$\varphi_{t+1}^{\alpha}$=$-U_{t+1}^{\alpha}-\gamma\text{log}\big(\! \sum_{\nu\in\mathcal{A}}\text{exp}\big( \frac{-\varphi_{t+2}^{\nu}}{\gamma}\big)\underline{\Gamma}\text{exp}\big( \frac{-\hat{\zeta}}{2(\underline{\Gamma})^2}\!\big)\! \big)$\\
 \bottomrule
 \end{tabular}
\end{table*}

\begin{figure}[!t]
\centering
\includegraphics[trim={1.6cm 0 1.6cm 0},width=\columnwidth]{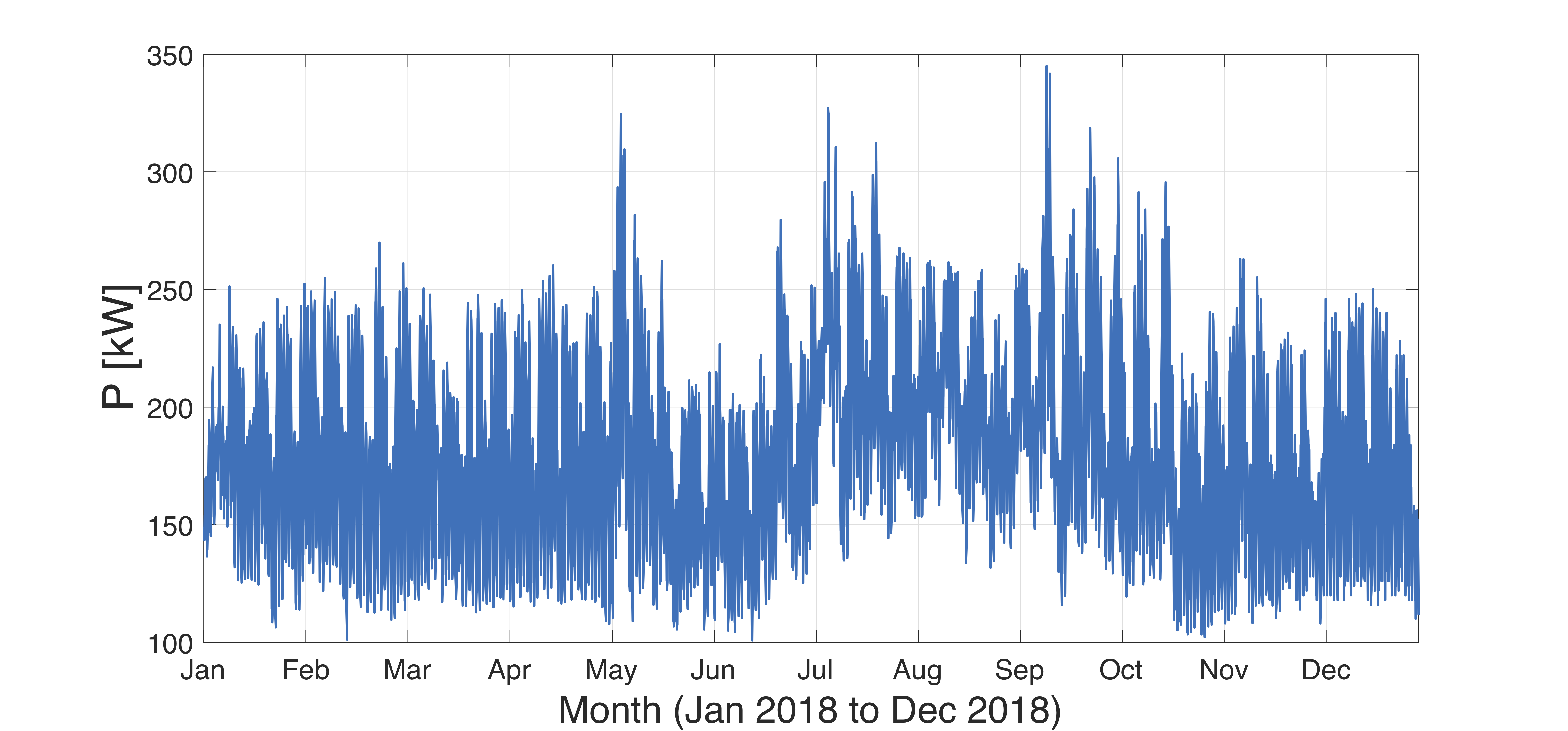}
\caption{One-year power consumption data for NYU Building 12.}
\label{fig:power_nyu}
\end{figure}

\begin{figure}[!t]
\centering
\subfloat[]{\includegraphics[height=3.8cm,width=0.45\columnwidth]{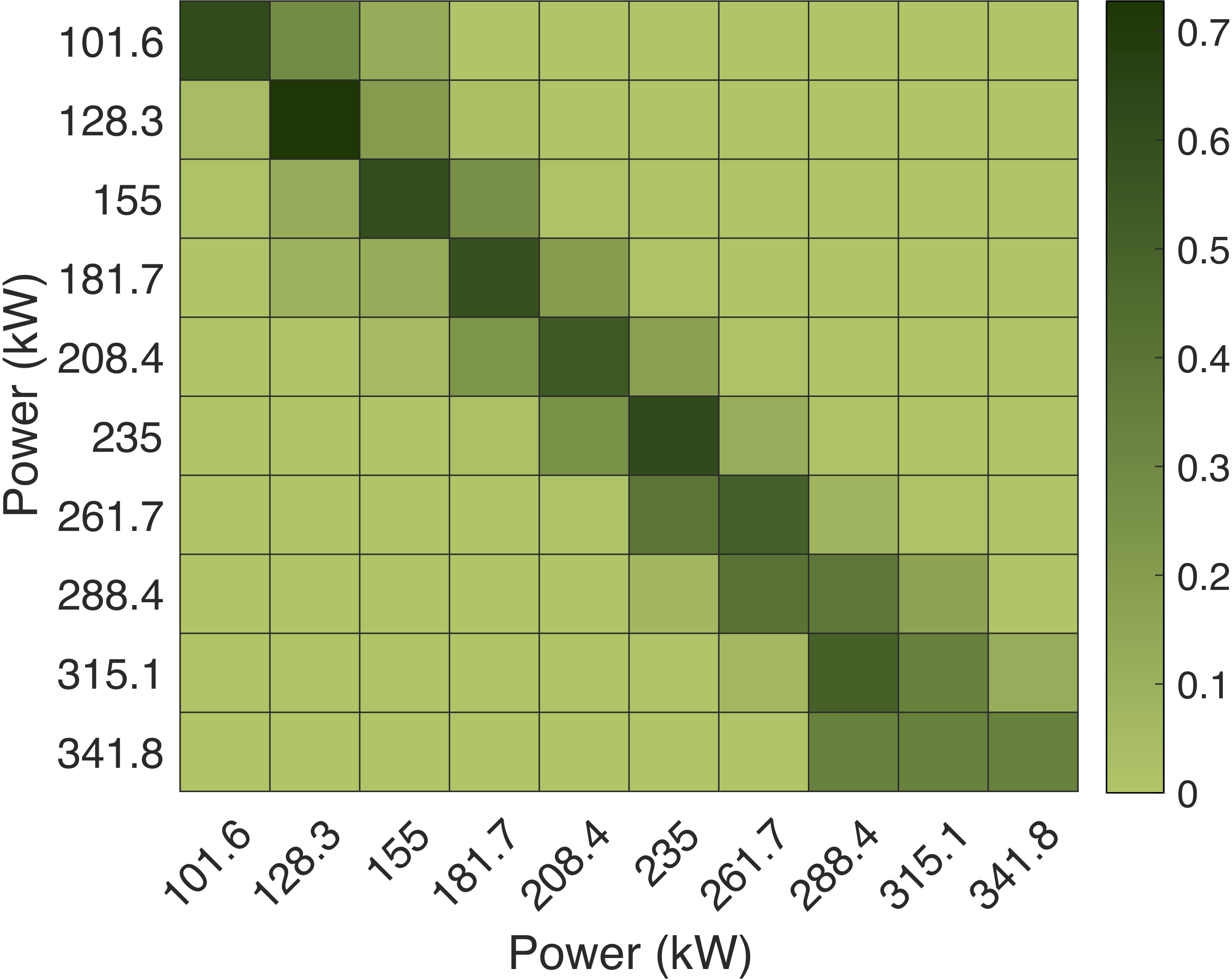}\ \ }
\subfloat[]{\includegraphics[height=3.6cm,width=0.45\columnwidth]{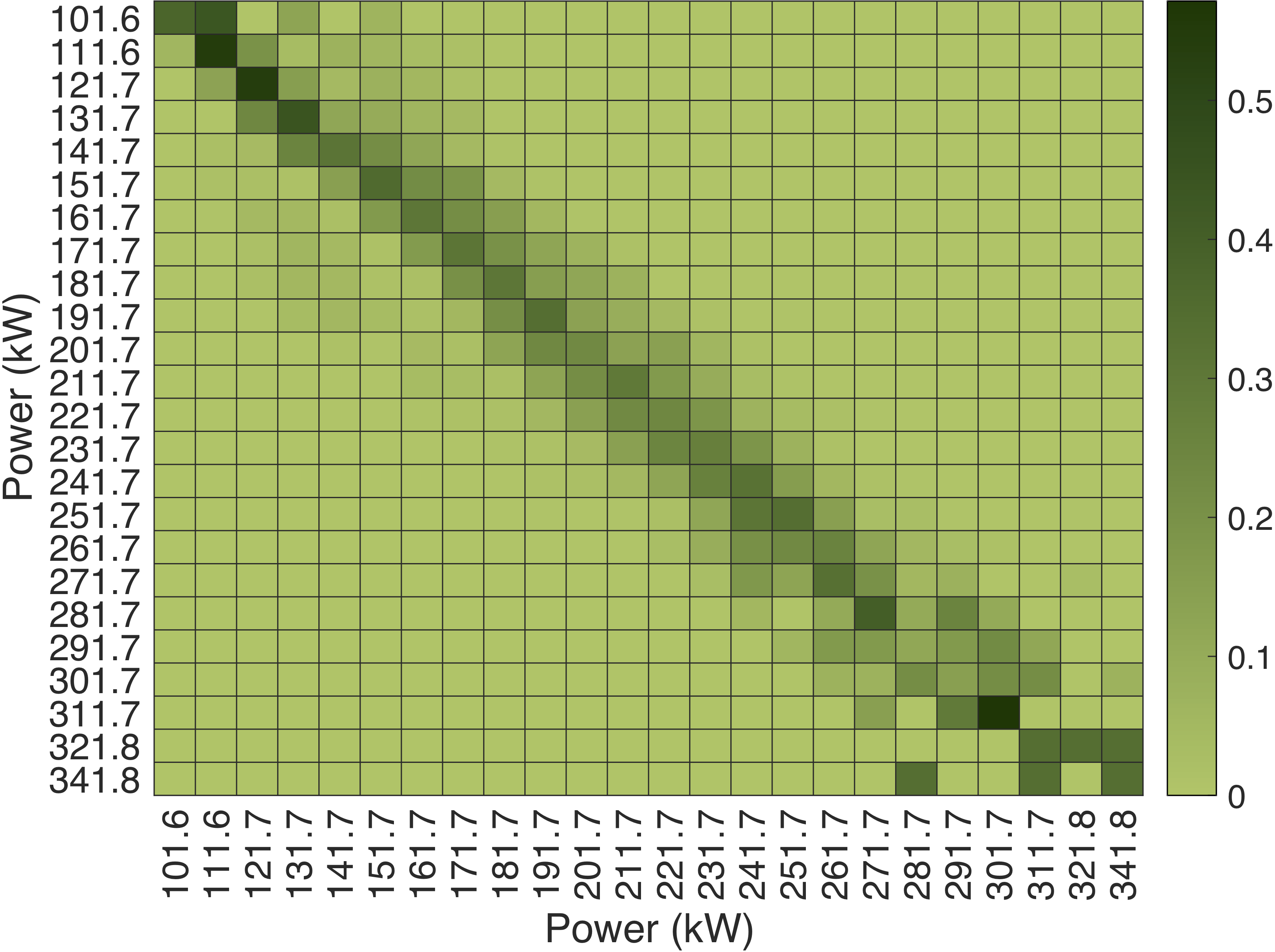}}
\caption{Transition probability matrix for NYU building 12 with (a) 10 states, and (b) 25 states.}
\label{fig:mp_nyu}
\end{figure}

Similarly to the application of the three-step procedure described in Section~\ref{sec:markov}-\ref{sec:mp_construction} and illustrated in Fig.~\ref{fig:mdp_three_steps} to residential households as explained above, we demonstrate the ability to construct the MP for a commercial building in an urban environment. As a demonstration site, we use the NYU campus building, abbreviated below as Building \#12, which is located in Manhattan, NY. This eight-story building has $\approx$70,000 square feet of mostly classrooms, meeting rooms and faculty offices. The building also has a basement, which is mainly used for storage. The HVAC system of the building consists of one Roof Top Unit (RTU), two chillers, two Air Handling Units (AHUs), two Fan Coil Units (FCUs), and 79 Variable Air Volume boxes (VAVs). The electric power consumption data is available for this building since 2016 and we display a year-long sample used to construct the MP in Fig.~\ref{fig:power_nyu}. The year-long pattern in Fig.~\ref{fig:power_nyu} is then used to develop the MP with 10 and 25 states, which are illustrated in Fig.~\ref{fig:mp_nyu}. Relative to the residential households in Figs. \ref{fig:mp_roman_case1}-\ref{fig:mp_roman_case2}, we observe that the case study with the single NYU building produces more diagonal matrices (see Fig.~\ref{fig:mp_nyu}), which means that the building can be gradually controlled from one power state to another, i.e. the net electric power consumption of this building is more on a par with affine control policies used for conventional generators.

Using the MPs for residential and commercial buildings constructed above, Section V will develop a suitable MDP to optimally control the built environments in multi-energy infrastructure systems.

\section{Markov Decision Process} \label{sec:mdp}

The MDP optimization problem operating the TCL ensemble can be stated as:

\begin{align}
&\hspace{-5mm}\underset{\substack{\rho,\mathcal{P}}}{\text{min}} \sum_{t\in\mathcal{T}} O^A_{b,t} := \mathbb{E}_{\rho}
\sum_{t \in \mathcal{T}} \sum_{\alpha \in \mathcal{A}} \big(-U_{t+1}^{\alpha} + \sum_{\beta \in \mathcal{A}} \gamma \log\! \frac{\mathcal{P}_{t}^{\alpha\beta}}{\overline{\mathcal{P}}_{}^{\alpha\beta}}\big) \label{MDP:objective}
\end{align}
\begin{align}
\text{s.t.} \nonumber \\
&\rho_{t+1}^{\alpha} = \sum_{\beta \in \mathcal{A}} \mathcal{P}_{t}^{\alpha\beta} \rho_{t}^{\beta}, \ \ \forall \alpha \in \mathcal{A}, t \in \mathcal{T} \label{MDP_evolution} \\
&\sum_{\alpha \in \mathcal{A}} \mathcal{P}_{t}^{\alpha \beta} = 1, \ \ \forall \beta \in \mathcal{A}, t \in \mathcal{T} \label{MDP_integrality}
\end{align}

where $\rho \in \mathbb{R}^{n}$ with $n=|\mathcal{A}|$ describes the dynamic state of the TCL ensemble such that its probability in state $\beta$ at time $t$ is given by $\rho_{t}^{\beta}$ and probability in state $\alpha$ at time $t+1$ is given by $\rho_{t+1}^{\alpha}$. $\mathcal{A}=\{\alpha,\beta,\cdots\}$ is the set of all possible states, which is obtained by discretizing the range of power consumption for each TCL ensemble. $\mathcal{P}_{t}^{\alpha\beta}$ characterizes the probability of transitioning from a given state $\beta$ at time $t$ to a next state $\alpha$ at time $t+1$.
Eq.~\eqref{MDP:objective} represents the objective function of the DR aggregator that controls the TCL ensemble and aims to maximize its expected utility ($U_{t+1}^{\alpha}$) and to minimize the discomfort cost for the TCL ensemble. The discomfort cost is computed using the Kullback-Leibler (KL) divergence, weighted by parameter $\gamma$. This divergence penalizes the difference between the transition decisions made by the DR aggregator ($\mathcal{P}_{t}^{\alpha\beta}$) and the default transitions of the TCL ensemble\footnote{Note that although $\mathcal{\overline{P}}^{\alpha \beta}$ is defined as time-independent, one can model it as time-dependent if there is enough observation data to construct a multi-period MP. As more empirical data on the TCL dispatch is collected over time, the more temporal fidelity can be achieved in representing default transitions.} ($\mathcal{\overline{P}}^{\alpha \beta}$), under the assumption that the latter represents first-choice preferences of TCL users. The choice of the Kullback-Leibler divergence for the penalty cost is motivated by its extensive use for modeling randomness of discrete and continuous time-series. Eq.~\eqref{MDP_evolution} describes the temporal evolution of the TCL ensemble from time $t$ to $t+1$ over time horizon $\mathcal{T}$. Eq.~\eqref{MDP_integrality} imposes the integrality constraint on the transition decisions optimized by the DR aggregator such that their total probability is equal to one. The active power ($p_{t}$) consumed by the TCL ensemble is computed using decisions $\rho^{\alpha}_{t}$ and rated active power $p^{\alpha}$ at each state as $p_{t}= \sum_{\alpha \in \mathcal{A}} p^{\alpha} \rho_{t}^{\alpha},\forall t \in \mathcal{T}$. Thus, the active power consumed can be controlled by means of optimizing decisions $\rho^{\alpha}_{t}$, which in turn can be parameterized using DR protocols of a given building (e.g., dimming certain lights, changing HVAC settings, or shutting down some of the elevators in the building).

The MDP optimization formulated in \eqref{MDP:objective}-\eqref{MDP_integrality} is a LS-MDP as introduced in \cite{LSMDP_todorov,LSMDP_Dvijotham,LS_optimal_control} and discussed in \cite{Chertkov_MDP,TCL_networks_Misha,Roman_MDP,Hassan_TCL}. LS-MDP problems can be solved analytically and the optimal policy derived from the LS-MDP is not a mapping of states to action variables, as in a conventional MDP, but is a mapping of a current state into a next-state distribution, which minimizes the expected next-state costs and the divergence cost between the default and controlled probability distributions. {The LS-MDP in \eqref{MDP:objective}-\eqref{MDP_integrality} can be efficiently solved at scale using dynamic programming. As shown in \cite[Appendix 1.9]{Chertkov_MDP}, the optimal policy obtained from \eqref{MDP:objective}-\eqref{MDP_integrality} can be expressed as ($\mathcal{P}_{t}^{\alpha \beta} = \frac{\overline{\mathcal{P}}^{\alpha \beta}z_{t+1}^{\alpha}}{\sum_{\alpha \in \mathcal{A}}\overline{\mathcal{P}}^{\alpha \beta}z_{t+1}^{\alpha}}$), as shown in Table~\ref{table_comparison}, where the value function $\varphi_{t+1}^{\alpha}$ for state $\alpha$ at time $t+1$ is encoded in the control through
\begin{align}
 z_{t+1}^{\alpha} = \text{exp}(-\varphi_{t+1}^{\alpha}/\gamma).\label{value_fn}
\end{align} We refer interested readers to \cite[Appendix 1.9]{Chertkov_MDP} for more details on the derivation.}

\subsection{Uncertainty Management} \label{sec:stochastic}
The default transition probabilities in the standard MDP formulation in Eqs.~\eqref{MDP:objective}-\eqref{MDP_integrality} are assumed to be perfectly known, which does not hold in real-world applications, where the TCL ensemble is subject to unknown external influences and uncertain human behavior. We model this parameter uncertainty by representing default transition probabilities $\overline{\mathcal{P}}^{\alpha\beta}$ as random variables $\overline{\boldsymbol{\mathcal{P}}}^{\alpha\beta}$ and assume that $\overline{\boldsymbol{\mathcal{P}}}^{\alpha\beta}$ follows a normal distribution with perfectly known mean $\overline{\mathcal{P}}^{\alpha\beta}$ and variance $\sigma^{2}$, i.e. $\overline{\boldsymbol{\mathcal{P}}}^{\alpha\beta}\sim N(\overline{\mathcal{P}}^{\alpha\beta},\sigma^{2})$. Under this fairly mild assumption on the uncertainty of $\overline{\boldsymbol{\mathcal{P}}}^{\alpha\beta}$, we can still use our prior work in \cite{MDP_Uncertainty_Ali} to derive the analytical optimal control policy shown in Table~\ref{table_comparison}. Relative to the optimal policy in the standard MDP, the stochastic extension differs in the term $\text{exp}\Big(\frac{-\sigma^2}{2({\overline{\mathcal{P}}^{\alpha\beta}})^2}\Big)$, which internalizes the uncertainty on uncontrolled transition probabilities into the stochastic control policy.

Since the parameters of the uncertainty distribution are not exactly known and are informed by a finite number of historical observations, the distribution is ambiguous over the available data. Hence, assuming a single distribution as in Section~\ref{sec:mdp}-\ref{sec:stochastic} may not capture the actual uncertainty on the MP matrix accurately. To overcome this limitation, we define the ambiguity set as $\mathbb{D}= [ \underline{\Gamma}\! \leq \overline{\mathcal{P}}^{\alpha\beta}\!\!\! \leq \overline{\Gamma} ,\underline{\hat{\zeta}}\! \leq {\sigma}^2\! \leq \overline{\hat{\zeta}}]$, where $\underline{\Gamma}$, $\overline{\Gamma}$, $\underline{\hat{\zeta}}$ and $\overline{\hat{\zeta}}$ are confidence bounds on the empirical mean and variance. Since $\overline{\mathcal{P}}^{\alpha\beta}$ and ${\sigma}^2$ can be respectively modeled by $t$- and Chi-Square ($\mathcal{X}^2$) distributions \cite{chi_sq_dist}, we compute these bounds as:
\begin{align}
& \!\!\!\underline{\Gamma} = \overline{\mathcal{P}}^{\alpha\beta}\!\! - t_{(1-\varsigma/2)} \frac{\hat{\sigma}}{\sqrt{N}} \text{ and } \ \overline{\Gamma} = \overline{\mathcal{P}}^{\alpha\beta}\!\! + t_{(1-\varsigma/2)} \frac{\hat{\sigma}}{\sqrt{N}}, \label{eq:ambiguity_set_mean}\\
& \underline{\hat{\zeta}} = \frac{(N-1)\hat{\sigma}^{2}}{\mathcal{X}^2_{(1-\xi)/2}} \text{ and } \ \overline{\hat{\zeta}} = \frac{(N-1)\hat{\sigma}^{2}}{\mathcal{X}^2_{\xi/2}}, \label{eq:ambiguity_set_var}
\end{align}
where $\xi$ and $\varsigma$ are confidence parameters on the bounds. As derived in our prior work in \cite{MDP_Uncertainty_Ali}, in this robust case we can also derive an analytical optimal control policy as given in Table~\ref{table_comparison}. This robust extension internalizes the information about set $\mathbb{D}$ and immunizes the optimal control policy for the worst-case realization of distribution parameters drawn from this set.
\subsection{Coupling the MDP and infrastructure constraints}
Although the standard, stochastic and robust MDP optimizations for dispatching the TCL ensembles in a given built environment allow to compute the flexibility that can be extracted for infrastructure operations, the MDP does not account for the deliver ability of this flexibility given network constraints. Hence, we propose to design an integrated optimization problem that includes the MDP and network flows. For the sake of illustration, we consider the coordination between the MDP and electric power distribution model, which can be modeled using the optimal power flow (OPF) framework.

Among multiple techniques available to implement an OPF problem, we select the chance-constrained OPF (CC-OPF) because it can accommodate AC power flows accurately and can robustly treat uncertain behavior imposed by volatile demand and photovoltaic resources. The choice of the CC-OPF over other methods is motivated by several advantages. First, chance constraints do not require discretizing a probability space, as required for scenario-based stochastic programming methods, and internalize continuous probability distributions of uncertain parameters \cite{Mieth_CC}. Additionally, chance constraints can be enforced over an ambiguity set, that can better fit non-Gaussian empirical data, \cite{Lubin_Yury_CC,Yury_Uncertainty_Sets,Roald_CC}. Furthermore, the CC-OPF scales well for large networks, especially relative to scenario-based stochastic programming \cite{Lubin_Yury_CC}, and can be implemented in a decentralized manner, \cite{Hassan_ADMM,Emiliano_CC}. Finally, recent studies demonstrate that chance constraints are well-suited for electricity pricing under uncertainty at the wholesale and distribution levels due to their ability to ensure market design properties (e.g., revenue adequacy, cost recovery and incentive compatibility),\cite{Dvorkin_CC}.

The integrated MDP and CC-OPF is formulated as:
\begin{align}
\begin{split}
&\underset{\substack{\rho,\mathcal{P},\Theta,p,q}}{\text{min}}
\sum_{t \in \mathcal{T}} \Bigg[ \sum_{b \in \mathcal{N}^{\mathcal{T}}} \underbrace{ O^A_{b,t} }_{\text{Aggregator}} + \Lambda_t \underbrace{O^D_t}_{\text{Utility}} \Bigg] \label{obj}
\end{split}\\
&\hspace{-11mm} \text{s.t.\qquad Eqs.}~\eqref{MDP_evolution}-\eqref{MDP_integrality} \label{int_mdp}\\
& p_{t,b}= \sum_{\alpha \in \mathcal{A}} p_b^{\alpha} \rho_{t,b}^{\alpha}, \quad \forall t \in \mathcal{T}, b \in \mathcal{N} \label{mdp_injP1} \\
& q_{t,b} = \sum_{\alpha \in \mathcal{A}} q_b^{\alpha} \rho_{t,b}^{\alpha} , \quad \forall t \in \mathcal{T}, b \in \mathcal{N} \label{mdp_injQ1} \\
& \text{CC-OPF Constraints, see \cite[Eqs.(19)-(26)]{Hassan_TCL}}, \label{int_opf}
\end{align}
where $O_{b,t}^{A}$ and $O_{b,t}^{D}$ represent the objective function of the aggregator (e.g. see Eq.~\ref{MDP:objective}) and utility (e.g. power loss minimization as in \cite{Hassan_TCL}) and parameter $\Lambda_t$ is a tariff that monetizes the power losses. Note that the other choices of the objective function can be used in \eqref{obj}. For simplicity, all CC-OPF constraints are concentrated in (8)-(11), see \cite{Hassan_TCL} for details, and set $\Theta$ denotes CC-OPF decision variables. The active and reactive power injections
of the TCL ensemble at bus $b \in \mathcal{N}$ of the distribution system, where $\mathcal{N}$ is the set of all buses in the distribution system, are computed in \eqref{mdp_injP1} and \eqref{mdp_injQ1} based on the rated active ($p_{b}^{\alpha}$) and reactive ($q_{b}^{\alpha}$) power at state $\alpha$ and its probability $\rho_{t,b}^{\alpha}$.

To solve the integrated optimization problem above, we propose a decomposition-based algorithm that divides the optimization tasks between the utility and aggregator, i.e. replicates the decentralized decision-making structure and minimizes communication needs among them. The proposed decomposition is based on the dual decomposition algorithm, \cite{dual_decomposition}, that allows the integrated problem to be decomposed into two separate MDP and CC-OPF subproblems that are solved iteratively until convergence and, in particular, allows for using different solution techniques for each subproblem (e.g., the MDP and CC-OPF are solved using dynamic and SOC programming methods, respectively.) The algorithm inherits the properties of the dual decomposition, including convergence properties and the ability to deal with non-convex decisions. Notably, the decomposed algorithm makes it possible to achieve the separation between spatial
and temporal variables, which accelerates its iterative performance.
The decomposed problem is then solved as illustrated in Fig. \ref{fig:std2} and described below:
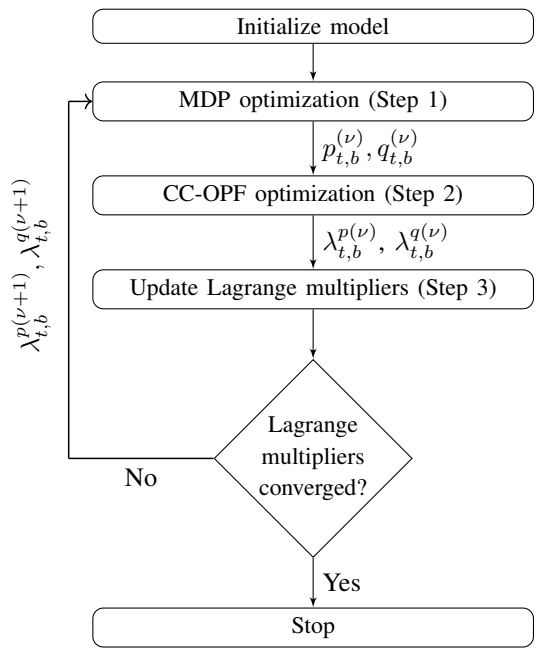
\begin{figure}[!t]
\centering
\tikzstyle{decision} = [diamond, draw, fill=white!10,
 text width=7.5em, text badly centered, node distance=1.50cm, inner sep=-8pt]
\tikzstyle{block} = [rectangle, draw, fill=white!10,
 text width=16em, text centered, rounded corners, node distance=1.25cm]
\tikzstyle{line} = [draw, -latex']
\tikzstyle{cloud} = [draw, ellipse,fill=red!20, node distance=2cm,
 minimum height=2em]

\begin{tikzpicture}[node distance = 1.25cm, auto]
 \node [block] (init) {\small Initialize model};
 \node [block, below of=init, node distance=1cm] (mdp_step1) {\small MDP optimization (Step 1)};
 \node [block, below of=mdp_step1] (ccopf_step2) {\small CC-OPF optimization (Step 2)};
 \node [block, below of=ccopf_step2, node distance=1.25cm] (Lagrange_step3) { \small Update Lagrange multipliers (Step 3)};
 \node [decision, below of=Lagrange_step3, node distance=2.25cm] (decide) { \small Lagrange multipliers converged?};
 \node [block, below of=decide, node distance=2.25cm] (stop) {\small Stop};
 \coordinate[left of=decide, xshift=-2.0cm] (c1);
 \coordinate[left of=mdp_step1, xshift=-2.0cm] (c2);
 \path [line] (init) -- (mdp_step1);
 \path [line] (mdp_step1) -- node {$p_{t,b}^{(\nu)}, q_{t,b}^{(\nu)}$} (ccopf_step2);
 \path [line] (ccopf_step2) -- node{$\lambda_{t,b}^{p(\nu)}$, $\lambda_{t,b}^{q(\nu)}$}(Lagrange_step3);
 \path [line] (Lagrange_step3) -- (decide);
 \path [line] (decide) -- node {Yes}(stop);
 \draw [-,thick, line width=0.2mm] (decide) -- (c1) node [midway] {No};
 \draw [-,thick, line width=0.2mm] (c1) -- (c2) node [midway, rotate=90, xshift=1.5cm, yshift=+0.5cm] {$\lambda_{t,b}^{p(\nu+1)}, \lambda_{t,b}^{q(\nu+1)}$};
 \draw [->,thick, line width=0.2mm] (c2) -- (mdp_step1);
\end{tikzpicture}
\vspace{5pt}
\caption{An iterative approach to decompose and solve the integrated MDP and CC-OPF optimization.}
\label{fig:std2}
\end{figure}

\begin{enumerate}
 \item Solve the following MDP optimization for each TCL ensemble to determine the optimal TCL dispatch:
 \begin{align}
 \forall b \in \mathcal{N}: \ & \underset{\substack{\rho,\mathcal{P}}}{\text{min}}
 \sum_{t \in \mathcal{T}} O_{b,t}^{A(\nu)} \label{eq:MDP_decompos1} \\
 &\hspace{-10mm} \text{Eqs.}~\eqref{MDP_evolution}-\eqref{MDP_integrality},~\eqref{mdp_injP1}-\eqref{mdp_injQ1} \label{eq:MDP_decompos2} \\
 &\hspace{-15mm}U_{t+1,b}^{\alpha(\nu)}\!\! =\! U_{t,b}^{\alpha(\nu)} \!+\! \lambda_{t,b}^{p(\nu)} p_b^{\alpha}\! + \!\lambda_{t,b}^{q(\nu)} q_b^{\alpha},\forall \alpha \!\in \!\mathcal{A},\! t\! \in \mathcal{T} \label{eq:MDP_decompos3}
 \end{align}
 where $\nu$ is an iteration counter and $\lambda_{t,b}^{p(\nu)}$ and $\lambda_{t,b}^{q(\nu)}$ are the Lagrange multipliers of Eq.~\eqref{mdp_injP1} and \eqref{mdp_injQ1}, respectively, obtained at the previous iteration of the algorithm.
 Hence, \mbox{$\lambda_{t,b}^{p(\nu=1)}=\lambda_{t,b}^{q(\nu=1)}=0$} during the first iteration. Note that while Eq.~\eqref{eq:MDP_decompos1}-\eqref{eq:MDP_decompos2} solve the original MDP problem as presented above, Eq.~\eqref{eq:MDP_decompos3} updates the next-step utility function of the aggregator, see Eq.~\eqref{MDP:objective}, using the most recently updated values of Lagrange multipliers $\lambda_{t,b}^{p(\nu)}$ and $\lambda_{t,b}^{q(\nu)}$. The optimization in \eqref{eq:MDP_decompos1}-\eqref{eq:MDP_decompos3} can be solved using either traditional dynamic programming methods (e.g., backward-forward algorithm that we used in \cite{Hassan_TCL}) or derived policies presented in Table \ref{table_comparison}.

 \item Solve the CC-OPF problem, where each TCL ensemble is parameterized using the values of Lagrange multipliers $\lambda_{t,b}^{p(\nu)}$:
 \begin{align}
 \begin{split}
 \hspace{-5mm}\forall t \in \mathcal{T}: & \underset{\substack{\Theta}} \min \sum_{t \in \mathcal{T}} \Lambda_t O^{D(\nu)}_t - \sum_{b\in\mathcal{N}}(\lambda_{t,b}^{p(\nu)}p_{t,b}^{(\nu)} + \lambda_{t,b}^{q(\nu)}q_{t,b}^{(\nu)})
 \end{split}\\
 &\hspace{-10mm}\text{s.t. \quad}~\text{CC-OPF Constraints, see \cite[Eqs.(19)-(26)]{Hassan_TCL}},
 \end{align}
 where the CC-OPF problems for all time intervals are solved in parallel using off-the-shelf convex solvers (e.g., CPLEX, Gurobi, MOSEK). When all subproblems are solved, Step 2 produces the optimal dispatch decisions for the distribution system given the TCL dispatch optimized at Step 1. The Lagrange multipliers produced at Step 2 are then used to trade-off the TCL dispatch among the MDP and CC-OPF optimization using the dual update in Step 3.
 \item Update the Lagrange multipliers to seek consensus decisions among the MDP and CC-OPF optimization problems:
 \begin{align}
 &\lambda_{t,b}^{p(\nu+1)} \leftarrow \lambda_{t,b}^{p(\nu)} + \delta\bigg(\sum_{\alpha \in \mathcal{A}}p_b^{\alpha }\rho_{t,b}^{\alpha(\nu)} - p_{t,b}^{(\nu)}\bigg)\\
 &\lambda_{t,b}^{q(\nu+1)} \leftarrow \lambda_{t,b}^{} + \delta\bigg(\sum_{\alpha \in \mathcal{A}}q_b^{\alpha(\nu)}\rho_{t,b}^{\alpha(\nu)} - q^{(\nu)}_{t,b}\bigg)
 \end{align}
 where $\delta$ is an exogenous parameter that can be tuned to improve computational performance.
\end{enumerate}

\subsection{Application to Building \#12 and NYU microgrid }

\begin{figure}[!t]
\vspace{2mm}
 \centering
 \includegraphics[width=1\columnwidth, clip=true, trim= 0mm 0mm 0mm 0mm]{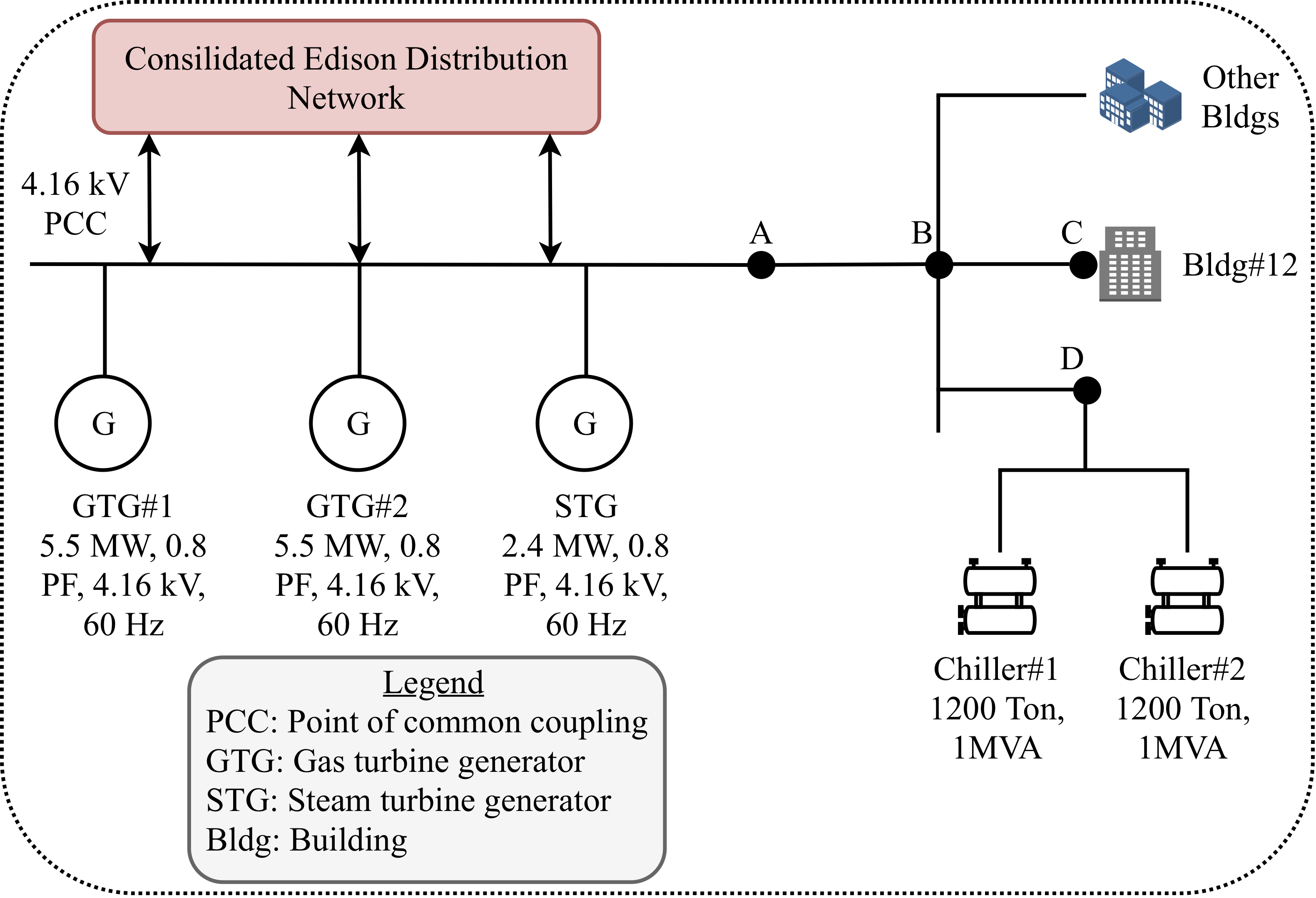}
 \caption{Schematic diagram of NYU microgrid depicting combined heat and power generation \cite{NYU_microgrid}, where AB, BC, and BD are distribution lines modeled in the CC-OPF implementation.}
 \label{fig:NYU_microgrid}
\end{figure}

Using the integrated MDP and CC-OPF optimization and decomposition algorithm described above, we analyze the dispatch flexibility of the NYU campus building \#12, which is a part of the NYU microgrid. Fig.~\ref{fig:NYU_microgrid} displays the NYU microgrid located in Manhattan, NY, which features one Combined Heat and Power (CHP) plant, supplying electricity to 22 buildings and heat to 37 buildings. We model this microgrid as a four-bus system (the buses are denoted in Fig.~\ref{fig:NYU_microgrid} by letters A, B, C and D), which are connected by distribution lines. The NYU microgrid is connected to ConEd at 4.16 kV point of common coupling (PCC) and can be islanded if needed. The islanding has been successfully demonstrated during Hurricane Sandy in 2012, when the microgrid islanded and continued to supply power and heat to critical NYU campus loads, \cite{NYU_sandy}. We implement the CC-OPF for the NYU microgrid under the assumption that it is voltage constrained and the allowed voltage fluctuation range is $\pm$5\% from nominal voltages at nodes A, B, C and D shown in Fig.~\ref{fig:NYU_microgrid}. Since this microgrid does not have flow constraints, we do not impose power flow limits in the CC-OPF implementation (which is typical for low-voltage distribution systems. The electricity is generated by two 5.5 MW (0.8 pf, 4.16 kV, 3-phase, 60 Hz) Gas Turbine Generators (GTG) and a 2.4 MW (0.8 pf, 4.16 kV, 3-phase, 60 Hz) Steam Turbine Generator (STG). The STG is driven by the steam generated by the hot exhaust produced as a by-product of gas turbines. After the steam is passed through the STG, it is used again to produce hot water for the campus in two high-temperature hot water exchangers and to operate chillers that provide cool water for air-conditioning. The microgrid has two 1 MVA chillers, which get electricity supply from both the CHP plant and ConEd's network.

\begin{figure}[!t]
\centering
\subfloat[]{\includegraphics[trim={0.2cm 0.2cm 0.2cm 0},width=0.9\columnwidth]{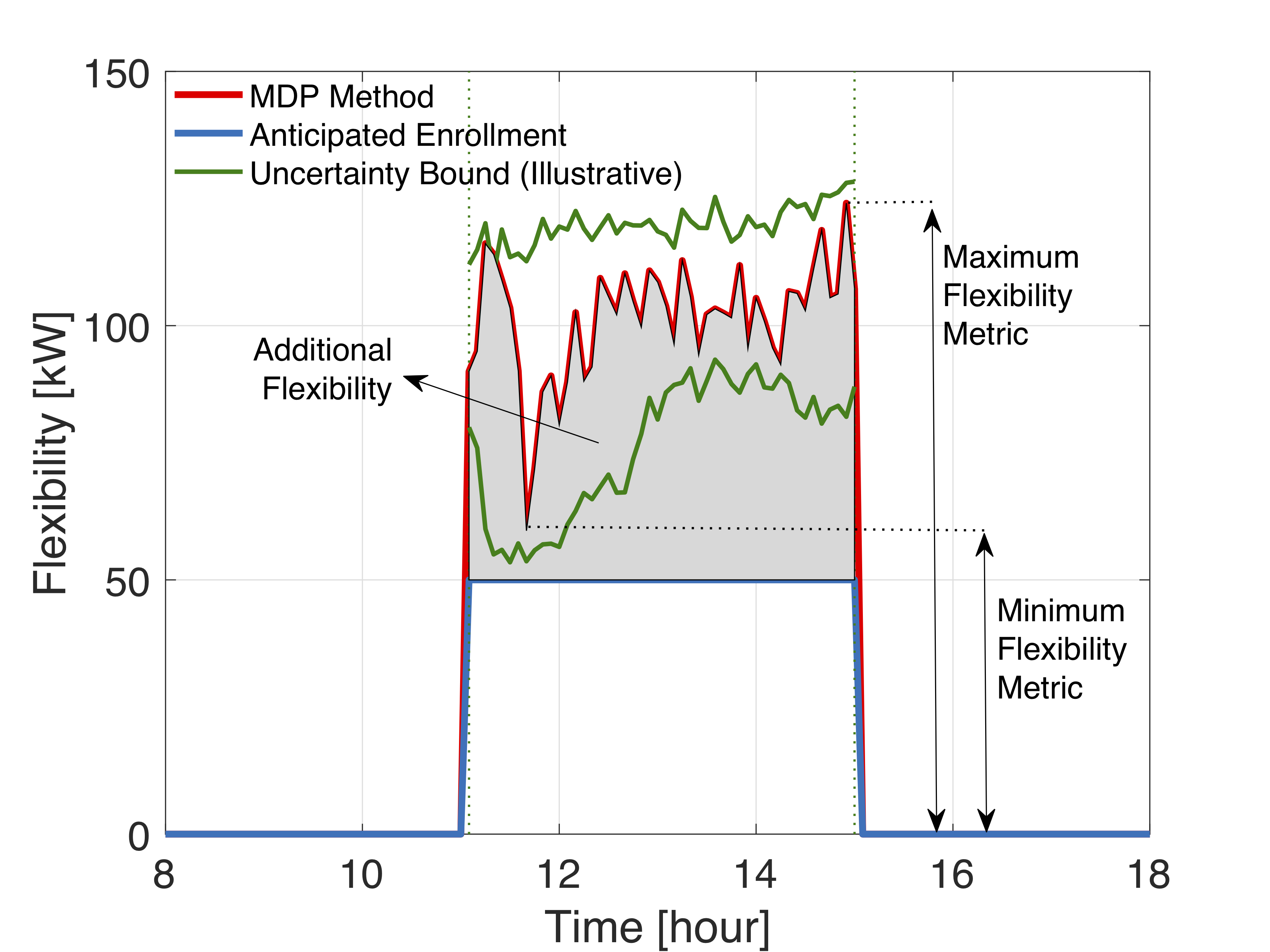}}\\
\vspace{-4mm}
\subfloat[]{\includegraphics[trim={0.2cm 0.2cm 0.2cm 0cm},width=0.9\columnwidth]{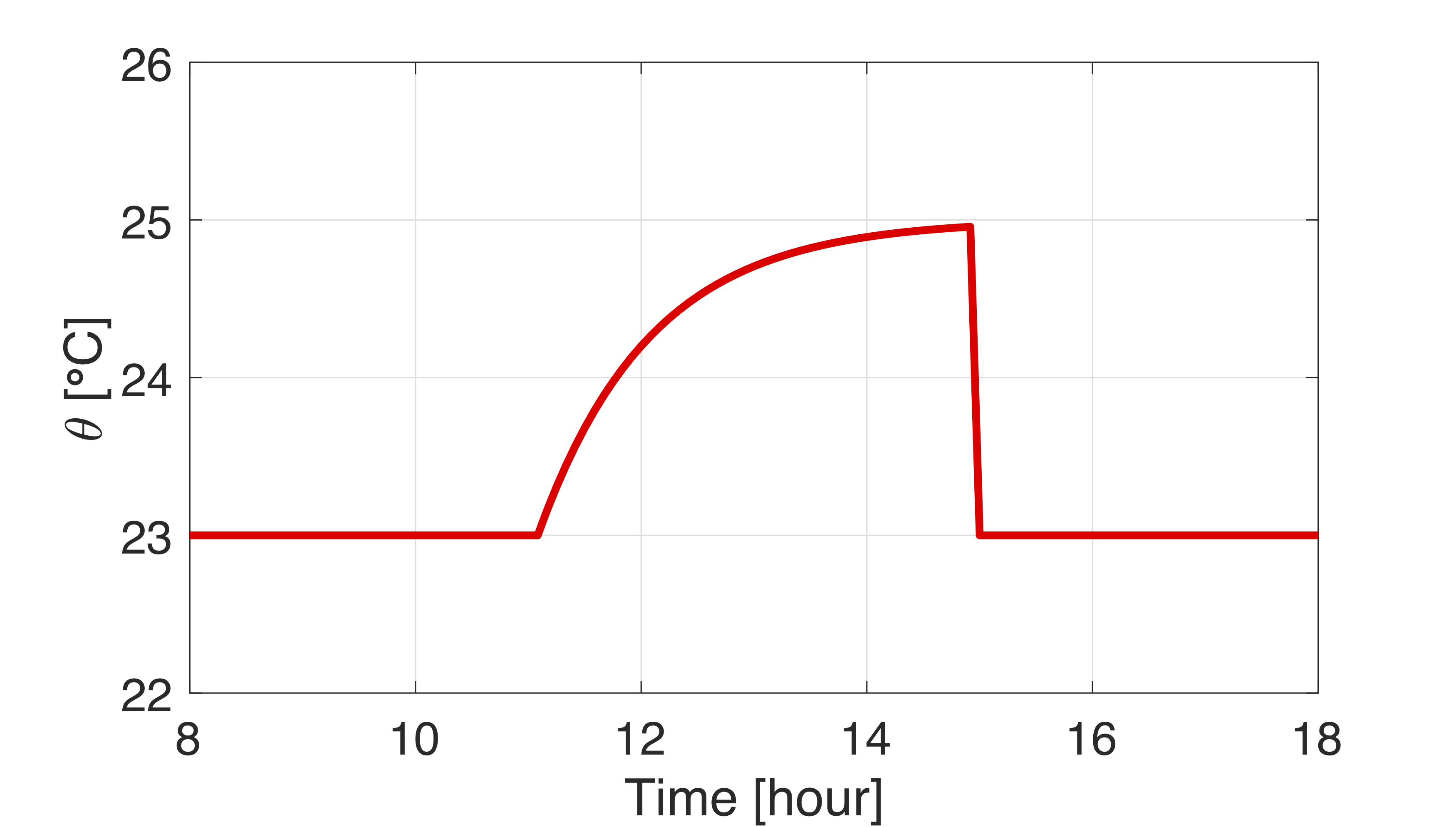}}
\caption{Demand response of NYU Building \#12 using the MDP method: (a) Simulation of the demand response event, (b) NYU Building \#12 indoor temperature profile.}
\label{fig:flexibility}
\end{figure}

\begin{figure}[!t]
\centering
\includegraphics[width=\columnwidth]{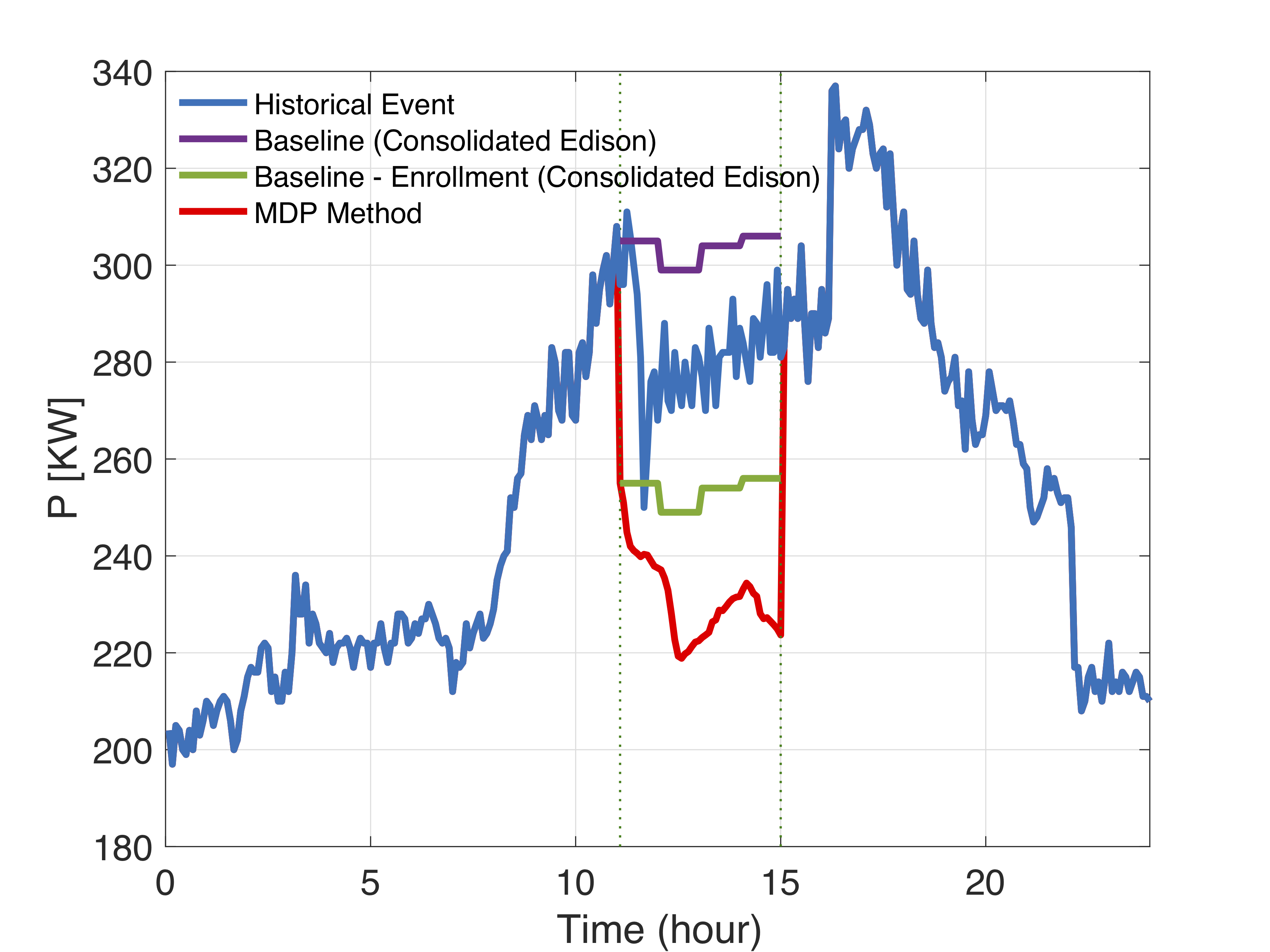}
\caption{Improvement in real-time demand response performance of NYU Building \#12. }
\label{fig:flexibility_realtime}
\end{figure}

Fig.~\ref{fig:flexibility}(a) demonstrates the improvement in extracting the demand-side flexibility from NYU Building \#12 using the proposed MDP approach as compared to the current practice used by ConEd for a historical demand response event that occurred from 11:00 through 15:00 hours on 11-26-2016. Based on the current practice, ConEd estimates the curtailment at 50 kW relative to the baseline of 299-306 kW. On the other hand, using the proposed MDP makes it possible to extract additional flexibility, which can be delivered via the NYU microgrid to the distribution network operated by ConEd. For contrast, we also display uncertainty limits on the demand response flexibility using the maximum and minimum response provided in other historical events. Notably, under the demand response participation decided by the MDP, the indoor temperature of the building is within a predefined comfort range as shown in Fig.~\ref{fig:flexibility}(b). Similarly, Fig. \ref{fig:flexibility_realtime} displays the improvement in the real-time performance of the NYU Building \#12 during the demand response event. As a reference, ConEd estimated a static enrollment of 50 kW during the DR event from the baseline of 299-306 kW\footnote{ConEd calculates the baseline by averaging the usage of each hourly interval of the top 5 days out of last 10 eligible weekdays (except the day before, holidays, low usage days - less than 25\% of the average usage level, and event days).}.

\section{Learning using MDPs}\label{sec:learning}
While the distributed optimization with central co-ordination for network-aware multi-energy systems is promising, it requires information of system parameters, aggregator's utility functions for optimal operation. Such information varies over different time-scales and may not be available in near-real time, due to non-ubiquitous metering equipment. For example, power distribution grids where controllable TCLs are located, have low observability over line-flows and structure. Similarly, consumer utility functions used in the MDP framework may not be appropriately known, unless through the use of consumer surveys etc. Recently, statistical learning of distribution grid state and parameters using real-time bus voltage measurements in the regime of partial observability has been discussed \cite{deka2020joint,sejunpscc}. In this section, we discuss active and passive approaches to efficiently estimate the utility function as well as optimal MDP policies using measurement data.

\subsection{Model-Free environment}
The MDP approach (see Eq.~\eqref{MDP:objective}) utilizes knowledge of the uncontrolled transition probability matrix $\overline{\mathcal{P}}^{\alpha\beta}$. While that may not be immediately available, by discretizing observed sample trajectories into bins, an empirical estimate $\overline{\mathcal{P}}^{\alpha\beta}_{emp}$ can be estimated. Interestingly, $\overline{\mathcal{P}}^{\alpha\beta}$ estimation can be by-passed and the optimal control directly estimated from samples. This is done by exploiting the specific structure in the optimal control law, described in Table \ref{table_comparison}. Plugging the control law and Eq.~\eqref{value_fn} into the MDP cost function, we get a fixed point equation for $z$, given below:

\begin{align}
&z^{\beta}_{t} = \text{exp}\Big(\frac{U_{t}^{\beta}}{\gamma}\Big) \sum_{\alpha}\overline{\mathcal{P}}^{\alpha \beta}z^{\alpha}_{t+1} \label{bellmen_reduced_3}
\end{align}
Using available samples, this can be solved by the following iterative procedure known as Z-learning \cite{LSMDP_todorov}, which is guaranteed to converge to the true solution for decaying weights $\eta_k$:
\begin{align}
&\hat{z}^{\beta}_{t,k} \leftarrow (1-\eta_k)\hat{z}^{\beta}_{t,k-1} + \eta_k \text{exp} \bigg(\frac{U^{\beta}_{t}}{\gamma} \bigg) \hat{z}^{\alpha}_{t+1,k-1}
\end{align}
It is worth noting that passive state transitions are sufficient to learn the control policy in Z-learning, and no user-intervention is required. Further the control law learnt can be robustified using the robust counterparts mentioned in Table \ref{table_comparison}.

\subsection{Customized Prices/Penalties}
While the Z-learning algorithm mentioned previously is able to learn the optimal control rule, it still depends on the parameter $\gamma$ that affects the cost associated with changing the transition probability. In a more general setting, $\gamma$ may vary from state to state and hence complicate the utility function associated with the aggregator. In such a setting, we advocate an `active' approach to estimate the associated cost function by calibrating the aggregator's response. For example, customized fluctuations in the active power cost $U$ may be used to identify the changes in aggregator response to identify which specific transitions are more flexible for affecting a change in the MDP control, and which are more restrictive. On a related setting, randomness in uncontrolled price signals and attempted actions can be internalized in the formulation as well to ensure price-robust operation of the MDP control following the learning step.
\section{Further extensions} \label{sec:future}
\subsection{Resiliency application}

Multi-energy infrastructure systems and built environments face natural and
anthropogenic (i.e., originating from human activity) cyber-physical threats that incur billions of
dollars in losses and casualties. A dramatic increase has been observed with \$2.9 trillion in direct
economic losses due to natural disasters (i.e., geo-physical and climate-related) during the past 20
years, with 77\% due to climate change. United States has suffered the greatest economic losses,
nearly \$1 trillion \cite{ipcc_2018}. Coastal states (e.g., New York, California, and Florida) are particularly
vulnerable to increased risk from natural disasters \cite{pelling2003vulnerability}. Cyber-attacks, on the other hand, not only
resulted in vital infrastructure failures \cite{6074981}, but also in data losses and privacy breaches \cite{Wheatley2016}. Cyber attacks on critical infrastructure systems are only increasing in frequency and severity as reported by the Center for Strategic \& International Studies \cite{cyber_csis}. For example, Kaspersky Security Network (KSN), an antivirus network company, revealed that 42.7\% of industrial control systems for multi-energy infrastructures that it protects were attacked in the year 2018 \cite{cyber_kaspersky}. 

The current practice of resiliency assessment on building and energy infrastructure is ad-hoc. It
fails to leverage or marginalize the ability of in-building resources to provide on-demand service
for alleviating stress and disturbances caused by extreme events. The uniqueness of the building
and energy infrastructure systems is that, unlike other aging infrastructure systems, they have a lot
of small-scale distributed resources that can be leveraged toward improving their joint resiliency. There is no consensus on a single metric or a set of metrics for defining and quantifying resiliency of multi-energy infrastructure systems that would internalize building dynamics. The proposed MDP framework can be leveraged to support building operations during extreme events, whether natural or
anthropogenic, and assist multi-energy infrastructures by providing back up flexibility to improve infrastructure preparedness and recovery.

\subsection{Beyond electricity network}
Aggregation of consumers into an ensemble can be extended to consumers drawing energy from the multi-energy system considered as a whole, i.e. including district heating and natural gas systems and their interdependencies with the power system. Furthermore, as consumer-end automation tools proliferate, it will become possible to design ensemble controls that account for the ability of consumers to switch between these three energy sources to satisfy their energy needs in the most efficient manner (e.g. least cost, energy conversation, environmentally responsible). From the multi-energy system perspective, a particular challenge will arise to account for the ability to use these consumer-end controls for storing and arbitraging energy between multi-energy systems. Among such consumer-end storage resources, heating, electric vehicles, gas-to-electricity equipment has been proven viable.

Although modeling such multi-energy consumers with the complex control and storage capabilities described above can be done similar to modeling a single-energy consumer in the proposed MDP framework, there are several technical gaps that need to be addressed. First, it will be required to identify parameters that characterize state and actions spaces of advanced multi-energy appliances. This, for example, should account not only for instant energy consumption (i.e. in terms of electricity, gas and heat) but also correlations among them and complex transitions from one energy source to another. This additional information should be internalized in the transition probability matrix and cost function to meet energy needs of consumers economically. Second, similarly to the integration of the MDP framework with the CC-OPF as described above, it will be useful to couple the MDP framework to decision support tools used for operating natural gas and heating systems, at the distribution and, possibly, transmission level. This multi-energy system optimization can, in turn, be extended to account for their inherent dynamics, including delays, mutual de-synchronization and temporal transients. For example, heat delivery (especially, in the modern setting of the third generation district heating systems) is a relatively slow process leading to significant delays. To mitigate these delays, one may take advantage of the ability of consumers to temporarily switch off their heat appliances and replace them with energy provided by the gas or electricity systems.

\subsection{MDP Enhancements}


The developed MDP framework is flexible and can be adopted as per the task at hand, with varying computational, communication, control, and cyber-security constraints. Depending on all of these requirements and, possibly others, the operator of the multi-energy system (or the aggregator of multi-energy resources) is expected to choose appropriate levels of `data resolution,' i.e. a number of units included in the ensemble, a number of states, a number of time steps and their duration, that will allow for  trading off performance and accuracy of the ensemble control. These choices are many and particulars will vary for different settings, optimization formulations and other considerations (e.g. engineering judgement or human-in-the-loop considerations).

In turn, the model design can also be hierarchical, i.e. one may use already available models of energy appliances or of their aggregation (e.g. in buildings or ensembles of buildings) to obtain a single model. For example, this single model can be a building ensemble within a given area (e.g. urban districts or rural enclaves). The designed model can in turn be used to collect necessary data for building a high-resolution MDP of the ensemble. This includes the reconstruction of cost function parameters and constrain definition for a given transition probability matrix of the ensemble subject to available control means. In turn, these control means are, in fact, heterogeneous. Hence, we also envision incorporating approaches from the domain of reinforcement learning to construct an accurate MDP from the finite and assorted data sets available (both energy measurements and existing DR protocols or instructions). Thus, the Z-learning approach discussed earlier should be considered among many other available options because it allows for simultaneously learning lacking measurements and limits on controllable actions. In particular, we envision using the proposed MDP and Z-learning frameworks in an online or sequential setting, where the exploration stage (i.e. data acquisition) and exploitation (i.e. solving the MDP using currently available data) are optimally balanced. One particular criterion for trading off the exploration and exploitation stages can be explored via joint risk measures that appropriately share techno-economic risk among different energy systems and between the system operator and energy customers.

\section{Conclusion}
This paper describes a modeling framework based on Markov Decision Processes to parameterize and model multi-energy dynamics in dispatch tools for multi-energy infrastructure systems. This framework makes it possible to convert and store energy of different types (electricity, heat, gas) at infrastructure edges, thus providing flexibility to infrastructure operators and planners. The paper discusses how this framework can be integrated with traditional modeling tools (e.g. energy management systems based on power, gas and heat flows) that are currently used in practice. We further discuss the application of the proposed framework from the viewpoint of traditional utilities, aggregators of small-scale energy resources, and consumers. Together with the proposed decentralized architecture, the proposed framework is generic, scalable and modular to accommodate emerging infrastructure-edge resources and communication technologies and can be used for various modeling and analysis tasks for optimally operating and designing future multi-energy systems.

\bibliographystyle{IEEEtran}
\bibliography{ref.bib,gas.bib,gasConrado.bib}

\end{document}